\documentclass[twocolumn]{aastex62}

\usepackage{verbatim}
\usepackage[usenames, dvipsnames]{}
\usepackage{multirow}

\newcommand{\Mearth}{\ensuremath{M_\earth}}
\newcommand{\Rearth}{\ensuremath{R_\earth}}
\newcommand{\rhoearth}{\ensuremath{\rho_\earth}}
\newcommand{\Rsun}{\ensuremath{R_\sun}}
\newcommand{\Msun}{\ensuremath{M_\sun}}
\newcommand{\Lsun}{\ensuremath{L_\sun}}
\newcommand{\Rstar}{\ensuremath{R_\star}}
\newcommand{\Mstar}{\ensuremath{M_\star}}

\newcommand{\Teffstar}{\ensuremath{T_{\rm eff}}}
\newcommand{\Lstar}{\ensuremath{L_\star}}
\newcommand{\Mpl}{\ensuremath{M_{pl}}}
\newcommand{\Rpl}{\ensuremath{R_{pl}}}
\newcommand{\Rhopl}{\ensuremath{\rho_{pl}}}

\newcommand{\ms}{\ensuremath{\rm m\,s^{-1}}}
\newcommand{\kms}{\ensuremath{\rm km\,s^{-1}}}

\newcommand{\masyr}{\ensuremath{\rm mas\,yr^{-1}}}

\newcommand{\pmra}{\ensuremath{\mu_{\alpha}}}
\newcommand{\pmdec}{\ensuremath{\mu_{\delta}}}

\newcommand{\Gaia}{{\it Gaia}}
\newcommand{\logg}{log\ensuremath{g}}

\newcommand{\teff}{\ensuremath{T_{\rm eff}}}
\newcommand{\Teff}{\ensuremath{T_{\rm eff}}}
\newcommand{\teq}{\ensuremath{T_{\rm eq}}}

\newcommand{\thisstar}{TOI-1685}

\newcommand{\TESS}{\emph{TESS}}
\newcommand{\HST}{\emph{HST}}
\newcommand{\JWST}{\emph{JWST}}
\newcommand{\VOSA}{\emph{VOSA}}

\newcommand{\ARIADNE}{\emph{ARIADNE}}

\newcommand{\logL}{\ensuremath{log(L/L_{\odot})}}

\newcommand{\fbol}{\ensuremath{f_{bol}}}
\newcommand{\mbol}{\ensuremath{m_{bol}}}
\newcommand{\Lbol}{\ensuremath{L_{bol}}}

\shorttitle{Updated parameters for TOI-1685 b}
\shortauthors{Burt et al. 2024}

\begin{document}

\title{TOI-1685 b is a Hot Rocky Super-Earth: \\ Updates to the Stellar and Planet Parameters of a Popular \JWST\ Cycle 2 Target\footnote{\textcopyright 2024 All rights reserved.}}

\author[0000-0002-0040-6815]{Jennifer~A.~Burt}
\affiliation{Jet Propulsion Laboratory, California Institute of Technology, 4800 Oak Grove Drive, Pasadena, CA 91109, USA}

\author[0000-0003-0030-332X]{Matthew J. Hooton}
\affiliation{Cavendish Laboratory, JJ Thomson Avenue, Cambridge CB3 0HE, UK}

\author[0000-0003-2008-1488]{Eric E. Mamajek}
\affiliation{Jet Propulsion Laboratory, California Institute of Technology, 4800 Oak Grove Drive, Pasadena, CA 91109, USA}

\author[0000-0003-0563-0493]{Oscar Barrag\'an}
\affiliation{Astrophysics, University of Oxford, Denys Wilkinson Building, Keble Road, Oxford, OX1 3RH, UK}

\author[0000-0003-3130-2282]{Sarah C. Millholland}
\affiliation{Department of Physics, Massachusetts Institute of Technology, Cambridge, MA 02139, USA; MIT Kavli Institute for Astrophysics and Space Research, Massachusetts Institute of Technology, Cambridge, MA 02139, USA}

\author[0000-0002-0692-7822]{Tyler R. Fairnington}
\affiliation{Centre for Astrophysics, University of Southern Queensland, West Street, Toowoomba, QLD 4350 Australia}

\author[0000-0003-0652-2902]{Chloe Fisher}
\affiliation{Astrophysics, University of Oxford, Denys Wilkinson Building, Keble Road, Oxford, OX1 3RH, UK}

\author[0000-0003-1312-9391]{Samuel P. Halverson}
\affiliation{Jet Propulsion Laboratory, California Institute of Technology, 4800 Oak Grove Drive, Pasadena, CA 91109, USA}

\author[0000-0003-0918-7484]{Chelsea X. Huang}
\affiliation{Centre for Astrophysics, University of Southern Queensland, West Street, Toowoomba, QLD 4350 Australia}

\author[0000-0003-2404-2427]{Madison Brady}
\affiliation{Department of Astronomy \& Astrophysics, University of Chicago, Chicago, IL, USA}

\author[0000-0003-4526-3747]{Andreas Seifahrt}
\affiliation{Department of Astronomy \& Astrophysics, University of Chicago, Chicago, IL, USA}

\author[0000-0002-5258-6846]{Eric Gaidos}
\affiliation{Department of Earth Sciences, University of Hawai'i at M$\bar{a}$noa, 1680 East-West Rd, Honolulu, HI 96822, USA}
\affiliation{Institute for Astrophysics, University of Vienna, T\"{u}rkenschanzstrasse 17, A-1180 Vienna, Austria}

\author[0000-0002-4671-2957]{Rafael Luque}
\affiliation{Department of Astronomy \& Astrophysics, University of Chicago, Chicago, IL, USA}

\author[0000-0003-0534-6388]{David Kasper}
\affiliation{Department of Astronomy \& Astrophysics, University of Chicago, Chicago, IL, USA}

\author[0000-0003-4733-6532]{Jacob L. Bean}
\affiliation{Department of Astronomy \& Astrophysics, University of Chicago, Chicago, IL, USA}

\begin{abstract}
We present an updated characterization of the TOI-1685 planetary system, which consists of a P$_{\rm{b}}$ = 0.69\,day USP super-Earth planet orbiting a nearby ($d$ = 37.6\,pc) M2.5V star (TIC 28900646, 2MASS J04342248+4302148). This planet was previously featured in two contemporaneous discovery papers, but the best-fit planet mass, radius, and bulk density values were discrepant allowing it to be interpreted either as a hot, bare rock or a 50\% H$_{2}$O / 50\% MgSiO$_{3}$ water world. TOI-1685 b will be observed in three independent JWST cycle two programs, two of which assume the planet is a water world while the third assumes that it is a hot rocky planet. Here we include a refined stellar classification with a focus on addressing the host star's metallicity, an updated planet radius measurement that includes two sectors of TESS data and multi-color photometry from a variety of ground-based facilities, and a more accurate dynamical mass measurement from a combined CARMENES, IRD, and MAROON-X radial velocity data set. We find that the star is very metal-rich ([Fe/H] $\simeq$ +0.3) and that the planet is systematically smaller, lower mass, and higher density than initially reported, with new best-fit parameters of \Rpl = 1.468 $^{+0.050}_{-0.051}$ \Rearth\ and \Mpl = 3.03$^{+0.33}_{-0.32}$ \Mearth. These results fall in between the previously derived values and suggest that TOI-1685 b is a hot, rocky, planet with an Earth-like density (\Rhopl = 5.3 $\pm$ 0.8 g cm$^{-3}$, or 0.96 \rhoearth), high equilibrium temperature (T$_{\rm{eq}}$ = 1062 $\pm$ 27 K) and negligible volatiles, rather than a water world.
\end{abstract} 

\keywords{
Exoplanets (498), 
Radial velocity (1332), 
Transit photometry (1709)}

\section{Introduction \label{sec:intro}}

To date, NASA's \TESS\ mission \citep{Ricker2015, Guerrero2021} has surveyed 93\% of the sky for at least 28 days and has detected over 1900 super-Earth and sub-Neptune sized planet candidates\footnote{Based on the ExoFOP TOI Catalog as of January 2024: https://exofop.ipac.caltech.edu/tess/view$\_$toi.php}. Many of these candidates are enticing targets for atmospheric follow up studies with missions such as the Hubble and James Webb Space Telescopes (\HST\ and \JWST, respectively) thanks to the relatively bright magnitudes of their host stars. Indeed, of the 52 unique planets selected for transit observations with \JWST\ in Cycle 2, 32 were first detected by \TESS\footnote{https://www.stsci.edu/jwst/science-execution/approved-programs/general-observers/cycle-2-go}.

Assessment of a planet's potential for such observations and the eventual interpretation of its atmospheric spectrum is reliant upon accurate and precise measurements of the planet's radius and mass which together dictate its scale height. As these properties are both obtained via indirect detection methods (generally transit photometry for the radius and radial velocity spectroscopy for the mass) accurate and precise knowledge of the host star's radius and mass is also crucial as these factor into the derived planet parameters.

Here we present an updated analysis and refined stellar and planetary parameters for one such small planet, TOI-1685 b. The planet was first published in a set of independent discovery papers: \citet{Bluhm2021} and \citet{Hirano2021}, hereafter denoted as B21 and H21, respectively. Both papers included photometry from \TESS\ sector 19 in their analyses, but the variety in the additional data sets incorporated and the analysis methods used between the two publications led to significant disagreements in their adopted values of stellar mass, stellar radius and planetary radius. Although the constraints on the radial velocity (RV) semi-amplitude and planet mass derived from the CARMENES RV data used in B21 and the IRD RV data used in H21 are consistent to within 1-$\sigma$, the corresponding planetary densities differ significantly due to different measurements of the planet's radius. B21 finds a best fit planet radius of \Rpl\ = 1.70 $\pm$ 0.07 \Rearth\, consistent with TOI-1685 b having a significant (50\%) water fraction, while H21 finds a best fit planet radius of \Rpl\ = 1.459 $\pm$ 0.065 \Rearth, which suggests the planet is a rocky super-Earth devoid of any significant water or volatile components. \citet{Luque2022} presented a joint analysis of the transit and RV data from both papers, adopting the stellar parameters from B21. Their derived system properties were in strong agreement with B21, but with a smaller best-fit mass which further favored a significant water contribution to this planet's composition. 

The proximity of the system, the planet's high planetary equilibrium temperature (\teq\ $\sim$ 1000K, B21 and H21) and the possibility of detecting atmospheric features in transmission and emission led to its inclusion in three successful \JWST\ programs \citep{Benneke_proposal,Luque_proposal,Fisher_proposal}, all of which are scheduled for 2024. Two of these proposals hold TOI-1685 b up as a bona fide water world, while the third assumes the planet is a rocky super-Earth and aims to study oxidation in its potential atmosphere.

We revisit the characterization of TOI-1685 and its planet, incorporating new precise RV measurements taken with the MAROON-X spectrograph, additional ground-based photometry from the MuSCAT network, and an additional sector of \TESS\ photometry. We include a study of the star's photometric metallicity, finding it to be metal-rich which results in a smaller stellar radius value than the previously published estimates. We highlight the expanded transit photometry and RV data sets, along with the broadband photometry used in our analysis in \S\ref{sec:data}, and then detail the updated stellar, transit, and RV analyses in \S\ref{sec:analysis}. We combine these results and present a refined interpretation of this system in \S\ref{sec:discussion}, touching on the planet's likely characteristics and its potential for atmospheric characterization during the scheduled \JWST\ programs, before concluding in \S5.


\section{Data \label{sec:data}}

\begin{table}[htbp]
\small
\caption{Astrometry and Photometry for TOI-1685}\label{tab:star}
\setlength{\tabcolsep}{2pt}
\begin{tabular}{lcc}
\hline\hline
Parameter & Value & Source \\
\hline \hline
TIC ID & TIC 28900646 & \citealt{Stassun2019}\\
2MASS ID & J04342248+4302148 & \citealt{Cutri2003}\\
Gaia DR3 ID & 252366608956186240 & Gaia DR3\\
RA\,(hh:mm:ss)\footnote{Gaia DR3 ICRS RA and Dec positions corrected to epoch J2000.0 via {\it Vizier}.} & 04:34:22.495 & Gaia DR3\\
Dec\,(dd:mm:ss) & +43:02:14.692 & Gaia DR3\\
\pmra\,(\masyr) & $37.762\pm0.022$ & Gaia DR3\\
\pmdec\,(\masyr) & $-87.062\pm0.018$ & Gaia DR3\\
$v_{rad}$ (\kms) & $-43.76\pm0.28$ & Gaia DR3\\
Parallax\,(mas) & $26.5893\pm0.0192$ & Gaia DR3\\
Distance\,(pc) & $37.596\pm0.022$ & \citealt{BailerJones2021}\\
SpType & M2.65V & \citealt{Terrien2015}\\
\hline
$B$ & 14.842 & Gaia DR3\\
$V$ & 13.362 & Gaia DR3\\
$R_c$ & 12.279 & Gaia DR3\\
$I_c$ & 10.928 & Gaia DR3\\
$G$ & $12.284570\pm0.002777$ & Gaia DR3\\
$T(TESS)$ & $11.112\pm0.007$ & \citealt{Stassun2019}\\
$J$ & $9.616\pm0.022$ & \citealt{Cutri2003}\\
$H$ & $9.005\pm0.023$ & \citealt{Cutri2003}\\
$K_{s}$ & $8.758\pm0.020$ & \citealt{Cutri2003}\\
\hline
$B - V$ (mag) & 1.480 & Gaia DR3\\
$G - K_{s}$ (mag) & $3.527\pm0.020$ & Gaia DR3 \&\\
  &  & \citealt{Cutri2003}\\
$B_p - R_p$ (mag) & 2.452011 & Gaia DR3\\
$B_p - G$ (mag) & 1.31245 & Gaia DR3\\
$G - R_p$ (mag) & 1.139555 & Gaia DR3\\
$M_V$ (mag) & 10.486 & This work\\
$M_G$ (mag) & $9.408\pm0.003$ & This work\\
$M_{Ks}$ (mag) & $5.882\pm0.020$ & This work\\
\hline
$U$\,(\kms) & $35.96\pm0.19$ & This work\footnote{Galactic Cartesian velocities (solar system barycentric frame): $U$ positive towards Galactic center, $V$ positive towards Galactic rotation, $W$ positive towards North Galactic Pole \citep[following ICRS to Galactic transformations from ][]{ESA1997}.}\\ 
$V$\,(\kms) & $-29.97\pm0.17$ & This work\\
$W$\,(\kms) & $-3.12\pm0.18$ & This work\\ 
\hline
\hline
\end{tabular}%
\end{table}

\subsection{Astrometry and Photometry \label{sec:astrom}}

\object{TOI-1685} is a magnitude $V=13.36$ early M dwarf star located in Perseus at a distance $d=37.596\pm0.022$ pc \citep[$\varpi$\, =\,$26.5893\pm0.0192$ mas;][]{GaiaDR3,BailerJones2021}. 
Astrometry and photometry for TOI-1685 are summarized in Table \ref{tab:star}. 
The star's position, proper motion, parallax, and \Gaia\, photometry are drawn from \Gaia\, Data Release 3 \citep{GaiaDR3}.
Optical photometry on the {\it Gaia} system ($G$, $G_{BP}$, and $G_{RP}$) 
and synthetic estimates of Johnson $B$ and $V$ and Cousins $R_{c}$ and $I_{c}$ are from \Gaia\, DR3, while the {\it TESS} synthetic magnitude $T$ is from the \TESS\ Input Catalog \citep{Stassun2019}. Near-infrared $J$, $H$, and $K_{s}$ photometry is adopted from 2MASS \citep{Cutri2003}.

\subsection{TESS Time Series Photometry \label{sec:TESS}}

TOI-1685 was observed by \TESS\ from UT 2019 November 27 to December 24 at two minute cadence as part of Sector 19 in \TESS\ primary mission, and then again from UT 2022 November 26 to December 23 at 20 second cadence as part of Guest Observer program 05064\footnote{``Impacts Of Superflares On \TESS\ Planetary Atmospheres'', PI: Ward Howard} in Sector 59.

The Science Processing Operations Center (SPOC) data \citep{Jenkins2016} for \thisstar\, available at the the Mikulski Archive for Space Telescopes (MAST) website\footnote{https://mast.stsci.edu} includes both simple aperture photometry (SAP) flux measurements \citep{Twicken2010, Morris2017} and presearch data conditioned simple aperture photometry (PDCSAP) flux measurements \citep{Smith2012, Stumpe2012, Stumpe2014}. The instrumental variations present in the SAP flux are removed in the PDC-SAP result. At the start of each orbit, thermal effects and strong scattered light impact the systematic error removal in PDC (see \TESS\ data release note DRN16 and DNR17). Before the fitting process described in \S\ref{sec:transit_phot}, we use the quality flags provided by SPOC to mask out unreliable segments of the time series. 

\subsection{Ground-based Time-Series Photometry \label{sec:ground}}

We endeavored to use all publicly available ground-based data of sufficient quality in our transit fit. We started by considering all of the ground based light curves presented in both B21 and H21, along with previously unpublished MuSCAT2 light curves from UT 2020 February 2 and 2021 February 2 \citep{Narita2019}. We discarded all of the MuSCAT2 light curves from UT 2021 January 19 due to poor weather conditions along with the $g'$- and $r'$-band MuSCAT2 light curves from UT 2021 January 29 and 2021 February 2 due to saturation of the best comparison star. We performed preliminary fits of the remaining light curves with all parameters except $R_\mathrm{p}/R_\star$ fixed to the median values measured from the \TESS\ light curves and discarded light curves where the measured $R_\mathrm{p}/R_\star$ was over 4-$\sigma$ from the \TESS\ value. These coincided with light curves particularly affected by time-correlated noise. The light curves that meet these criteria, and which we therefore use in the transit analysis described in \S\ref{sec:transit_phot}, are listed in Table \ref{tab:ground} and are available via Zenodo at this \dataset[DOI]{10.5281/zenodo.11105468}.

\begin{table}[ht]
\centering
\caption{Details of the ground-based transit light curves of TOI-1685 b that we included in the transit analyses presented in \S\ref{sec:transit_phot}, and the auxiliary parameters used in the detrending.}
\begin{tabular}{ccccc}
\hline\hline
Date & Instrument & Band & Detrending \\
\hline \vspace{2pt}
\multirow{3}{*}{2020-02-02} & \multirow{3}{*}{MuSCAT2$^{a}$} & $r'$ & \textit{am}, $\delta x$, $\delta y$, \textit{FWHM} \\
 & & $i'$ & \textit{am}, $\delta x$, $\delta y$, \textit{FWHM} \\
 & & $z_s$ & \textit{am}, $\delta x$, $\delta y$, \textit{FWHM} \\\hline
2020-03-08 & PESTO$^{b}$ & $i'$ & \textit{am}, \textit{bg} \\\hline
2020-11-07 & Sinistro$^{c}$ & $i'$ & \textit{am}, \textit{peak} \\\hline
2020-11-11 & Sinistro & $i'$ & \textit{am}, $\delta y$ \\\hline
2020-11-24 & MuSCAT$^{d}$ & $z_s$ & \textit{am}, $\delta x$, $\delta y$, \textit{FWHM}, \textit{peak} \\\hline
\multirow{3}{*}{2021-01-12} & \multirow{3}{*}{MuSCAT} & $g'$ & \textit{am}, $\delta x$, $\delta y$, \textit{FWHM}, \textit{peak} \\
 & & $r'$ & \textit{am}, $\delta x$, $\delta y$, \textit{FWHM}, \textit{peak} \\
 & & $z_s$ & \textit{am}, $\delta x$, $\delta y$, \textit{FWHM}, \textit{peak} \\\hline
\multirow{2}{*}{2021-01-14} & \multirow{2}{*}{MuSCAT} & $g'$ & \textit{am}, $\delta x$, $\delta y$, \textit{FWHM}, \textit{peak} \\
 &  & $z_s$ & \textit{am}, $\delta x$, $\delta y$, \textit{FWHM}, \textit{peak} \\\hline
\multirow{4}{*}{2021-02-01} & \multirow{4}{*}{MuSCAT3$^{e}$} & $g'$ & \textit{am}, $\delta x$, $\delta y$, \textit{FWHM}, \textit{peak} \\
 & & $r'$ & \textit{am}, $\delta x$, $\delta y$, \textit{FWHM}, \textit{peak} \\
 & & $i'$ & \textit{am}, $\delta x$, $\delta y$, \textit{FWHM}, \textit{peak} \\
 & & $z_s$ & \textit{am}, $\delta x$, $\delta y$, \textit{FWHM}, \textit{peak} \\\hline
\multirow{2}{*}{2021-02-02} & \multirow{2}{*}{MuSCAT2} & $i'$ & \textit{am}, $\delta x$, $\delta y$, \textit{FWHM} \\
 & & $z_s$ & \textit{am}, $\delta x$, $\delta y$, \textit{FWHM} \\\hline\hline

\end{tabular}
\label{tab:ground}
\tablecomments{Auxiliary parameter key: \textit{am} -- airmass, \textit{bg} -- background, $\delta x$ -- $x$ offset, $\delta y$ -- $y$ offset, \textit{FWHM} -- PSF full width at half maximum, \textit{peak} -- PSF peak brightness. \\
Instrument references: (a) \citet{Narita2019}, (b) https://omm-astro.ca/en/instrument/pesto/, (c) https://lco.global/observatory/instruments/sinistro/, (d) \citet{Narita2015}, (e) \citet{Narita2020}. }
\end{table}

MuSCAT light curves from UT 2020 January 24, 2021 January 10, 2021 January 12, and 2021 January 14 and MuSCAT3 light curves from UT 2021 February 1 were originally presented in H21. MuSCAT2 light curves from UT 2021 January 29 were originally presented in B21.
\subsection{Time Series Radial Velocities \label{sec:RV}}

TOI-1685 was included as a target during the 2020B and 2021B observing semesters on the MAROON-X spectrograph on the Gemini North telescope \citep{Seifahrt2016, Seifahrt2018, Seifahrt2020, Seifahrt2022}. MAROON-X is a stabilized, high-resolution (R $\approx$ 85,000), fiber-fed echelle spectrograph spanning 500 – 920 nm that was designed specifically for measuring precision radial velocities of M dwarf stars. The spectrograph employs two wavelength channels, one in the visible (500 – 670 nm, abbreviated as MX-Blue) and one in the NIR (650 – 900 nm, abbreviated as MX-Red). Both channels are exposed simultaneously when observing a target and each channel produces an independent velocity measurement.

A total of 17 MAROON-X epochs were obtained from November to December 2020, producing 34 velocity measurements (17 from the MX-Blue arm and 17 from the MX-Red arm). Another 10 MAROON-X epochs were obtained between October and November 2021, producing 20 velocity measurements (10 observations each from the MX-Blue and MX-Red arms). The MAROON-X velocities are extracted using a customized version of the SERVAL package \citep{Zechmeister2018} and have the following mean uncertainties: \\ MX-Blue 2020 : 1.66 \ms ; MX-Red 2020 : 1.02 \ms ; MX-Blue 2021 : 1.33 \ms ; MX-Red 2021 : 0.77 \ms.

The star was previously observed using the CARMENES spectrograph on the 3.5-m Calar Alto telescope \citep{Quirrenbach2014} and the IRD spectrograph on the Subaru telescope \citep{Kotani2018}. The results of these independent RV analyses were published in B21 and H21 for CARMENES and IRD, respectively. Following H21, we discard the IRD data taken on UT 2020 February 2, but all other observations from the three instruments are included in the RV analysis detailed in \S\ref{sec:rv_analysis}. The CARMENES data have a mean uncertainty of 2.44 \ms, while the IRD data has a mean uncertainty of 4.08 \ms. The time series RVs from all three instruments are presented in the Appendix, and the full RV dataset containing all outputs of the SERVAL pipeline is available on Zenodo at \dataset[DOI]{10.5281/zenodo.11105468}.


\section{Analysis \label{sec:analysis}}


\subsection{Stellar Parameter Analyses \label{sec:stellar_params}}

The two exoplanet discovery papers by B21 and H21 differ in the adopted values of the star's estimated metallicity, mass, and radius; all of which impact the conversion of a radial velocity semi-amplitude to a planetary mass and of a photometric transit depth to a planetary radius.
H21 estimated the star's metallicity to be [Fe/H] = $0.14\pm0.12$, its mass to be \Mstar\ = $0.460\pm0.011$ \Msun, and its radius to be \Rstar\ = $0.459\pm0.013$ \Rsun, relying on the parameters from the TIC \citep{Stassun2019} and calibration of \citet{Mann2015}.
B21 estimated the star's metallicity to be [Fe/H] = $-0.13\pm0.16$, its mass to be \Mstar\ = $0.495\pm0.019$ \Msun\, and its radius to be \Rstar\ = $0.492\pm0.015$ \Rsun\ using the M dwarf mass-radius calibration of \citet{Schweitzer2019}. 
We first discuss and try to improve upon the estimation of the star's metallicity before attempting to improve the star's radius and mass.\\

\subsubsection{Color-Magnitude Position and Spectral Type \label{sec:cmd}}

\begin{figure}
\centering
\includegraphics[width=.45\textwidth]{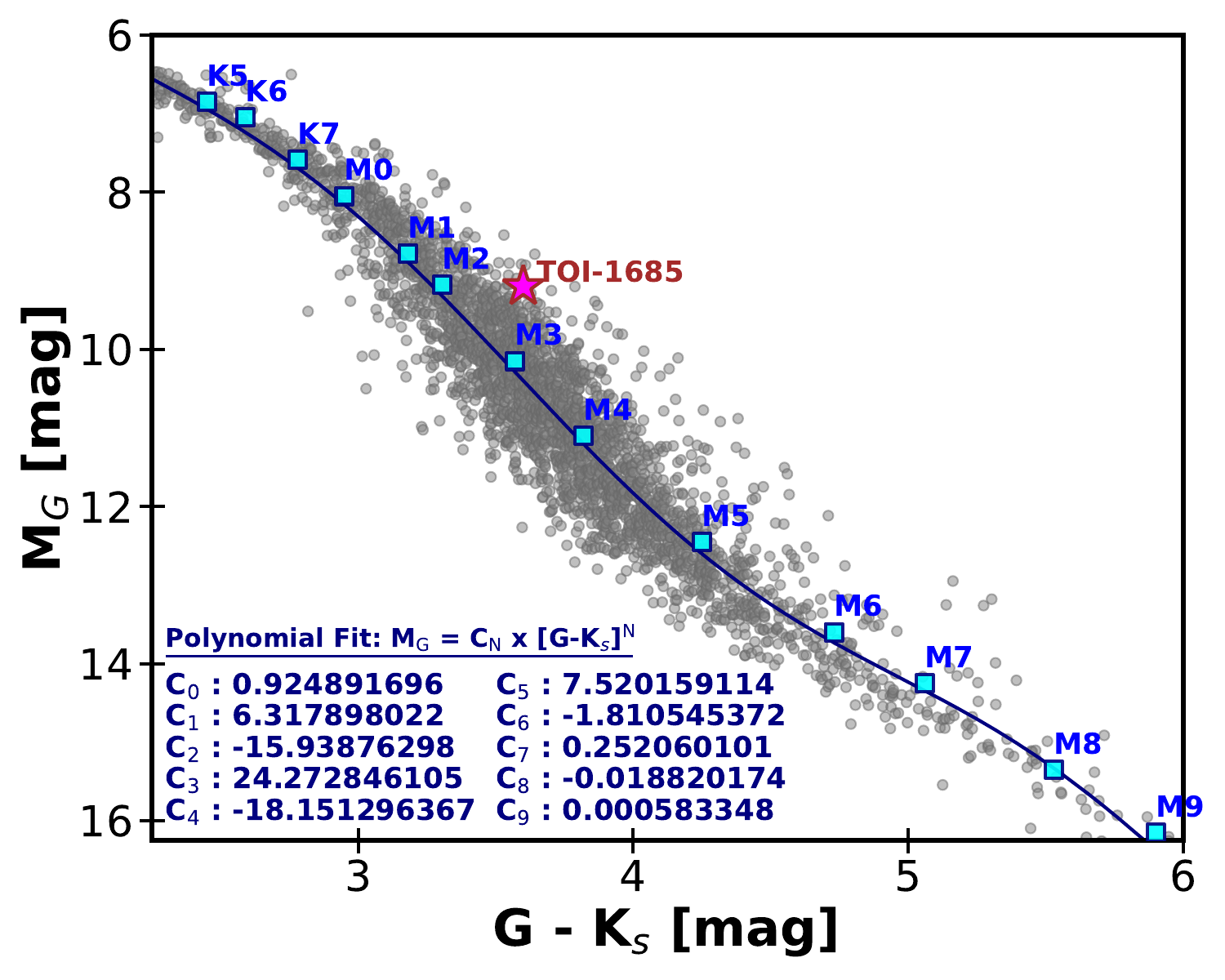}
\caption{Color-magnitude diagram ($G-K_{s}$ vs. $M_{G}$, with the $G$ magnitude taken from \citet{GaiaDR3} and the 2MASS $K_s$ magnitude from \citet{Cutri2003}) for stars within 25 pc ($\varpi$ $>$ 40\,mas) in SIMBAD, subtracting known white dwarfs, binary stars, and giants. The vast majority of the fiducial values for parallax $\varpi$ and magnitudes $G$ and $K_s$ in SIMBAD come from \Gaia\, DR3 for the first two and 2MASS for the latter. The blue line is a 9th order polynomial fit, with spectral types labeled at their mean colors from the updated online table from \cite{Pecaut2013}. TOI-1685, depicted as a pink star, is 0.57 mag brighter than the main sequence locus.
\label{fig:gk}}
\end{figure}

We plot the star's color-magnitude diagram position ($G-K_s$ vs. $M_G$) in Figure \ref{fig:gk}. For comparison, we plot the color-magnitude positions for nearby ($d$ $<$ 25\,pc; $\varpi$ $>$ 40\,mas) K and M dwarfs from SIMBAD. 
After removing known white dwarfs, evolved stars, and binaries from the SIMBAD sample, a 9th order polynomial is fit through the color-magnitude sequence. TOI-1685 sits 0.57 mag above this sequence ($\Delta M_{G}$ = -0.57) which requires an explanation. 

The null results of the High Resolution Imaging presented in the B21 and H21 papers, the lack of any long-term radial velocity trend, and the star's low \Gaia\, DR3 RUWE value (RUWE = 1.179) suggests that the star lacks an unresolved stellar companion\footnote{RUWE $>$ 1.4 is typically considered a signal of unresolved stellar multiplicity \citep{Lindegren2021}}, hinting that the star must be either metal-rich or pre-MS \citep[e.g.][]{Johnson2009,Bell2015} or be subject to significant stellar reddening. 

Stellar reddening can be ruled out as both the B21 and H21 spectroscopic analysis show that the star is indeed an early M dwarf whose effective temperature and color estimates are consistent, and at $d = 37.6$\,pc \citep{BailerJones2021} the star is well within the Local Bubble, where reddening should be negligible \citep{Reis2011,Lallement2018}. 
Further, there are no indications that the star could be pre-MS. TOI-1685 is slow-rotating \citep[$v$sin$i$ $<$ 2.0 \kms;][]{Marfil2021}, relatively inactive, and its relatively high space velocity ($S_{tot}$ = 47\,\kms) is not consistent with membership in any of $<$100\,pc young moving groups or clusters \citep[using BANYAN $\Sigma$;][]{Gagne2018}. 
Hence we interpret the star's overluminosity in color-magnitude space as an indicator of high metallicity.

The spectral subtype for TOI-1685 has been variously quoted as 
M1 \citep{Bai2018}, 
M2 \citep{Lu2019,Xiang2019}, 
M2.3$\pm$0.5 \citep{Sebastian2021}, 
M2.4 \citep{Birky2020},
M2.5 \citep{Zhong2015}, 
M2.65 \citep{Terrien2015}, 
M3.0 \citep{Zhong2015}, and 
M3.02 \citep{Terrien2015}. 
Through comparison of the broadband visible and near-infrared absolute magnitudes of TOI-1685 (those of \citealt{Kirkpatrick1991}) to M dwarf standard stars using the 2MASS photometry, \Gaia\, DR3 photometry, and astrometry compiled in SIMBAD \citep{Cutri2003,GaiaDR3,Wenger2000}, we find that TOI-1685 is most similar to three M2.5V standard stars: \\
TOI-1685 ($G-K_s$ = 3.53, $M_G$ = 9.41, $M_{Ks}$ = 5.88)\\
GJ 381\footnote{A tight binary ($\sim$0.2\arcsec) unresolved in \Gaia\, or 2MASS \citep{Mann2019}.} ($G-K_s$ = 3.42, $M_G$ = 9.30, $M_{Ks}$ = 5.88),\\
GJ 767B ($G-K_s$ = 3.65, $M_G$ = 9.41, $M_{Ks}$ = 5.76),\\
GJ 250B ($G-K_s$ = 3.37, $M_G$ = 9.39, $M_{Ks}$ = 6.01).
So the absolute photometry for TOI-1685 is at least consistent with a spectral type of approximately M2.5V, similar to that quoted by \citet{Terrien2015}, \citet{Zhong2015}, and \citet{Sebastian2021}.

\subsubsection{Stellar Metallicity \label{sec:metal}}

There are several published metallicity estimates for TOI-1685 which are summarized in Table \ref{tab:stellar_params}. These range from -0.22 to +0.22, with numerous entries in between. Safe to say, the metallicity for the star seems to be poorly constrained, other than that it is likely within a couple tenths of a dex of solar. We therefore proceed to generate an independent photometric metallicity estimate of this star.
\begin{table*}[ht]
\scriptsize
\caption{Stellar Parameter Estimates for TOI-1685}\label{tab:stellar_params}
\setlength{\tabcolsep}{2pt}
\begin{tabular}{lcrlcrlcr}
\hline\hline
\multicolumn{9}{c}{\textbf{Published Literature Values}} \\ \hline
\textbf{Param} & \textbf{Value}   & \multicolumn{1}{r|}{\textbf{Reference}} & \textbf{Param} & \textbf{Value}   & \multicolumn{1}{r|}{\textbf{Reference}} & \textbf{Param} & \textbf{Value} & \multicolumn{1}{r}{\textbf{Reference}} \\ \hline
Fe/H & $-0.13\pm0.16$   & \multicolumn{1}{r||}{B21}    & \Mstar\ & $0.495\pm0.019$  & \multicolumn{1}{r|}{B21}    & \Rstar\ & 0.492 $\pm$ 0.015   & B21    \\
Fe/H  & $+0.14\pm0.12$   & \multicolumn{1}{r||}{H21}    & \Mstar\ & $0.460\pm0.011$  & \multicolumn{1}{r|}{H21}    & \Rstar\ & 0.459 $\pm$ 0.013   & H21    \\
Fe/H & $-0.20$ & \multicolumn{1}{r|}{ \cite{Ding2022} (a)}  & \Mstar\ & $0.4600\pm0.020$ & \multicolumn{1}{r|}{\cite{Stassun2019}} & \Rstar\ & 0.462 $\pm$ 0.014   &  \cite{Stassun2019}    \\
Fe/H  & $-0.18$ & \multicolumn{1}{r|}{ \cite{Ding2022} (a)}  & \Mstar\ & $0.46\pm0.01$    & \multicolumn{1}{r|}{ \cite{Sebastian2021}}   & & -- & --    \\
Fe/H  & $-0.064\pm0.008$ & \multicolumn{1}{r|}{ \cite{Sprague2022}} & \Mstar\ & $0.461^{+0.003}_{-0.018}$ & \multicolumn{1}{r|}{ \cite{Morrell2019}} & & -- & --    \\
Fe/H  & $-0.05\pm0.05$   & \multicolumn{1}{r|}{ \cite{Yu2023}}  & \Mstar\ & $0.4999\pm0.00044$ & \multicolumn{1}{r|}{ \cite{Queiroz2020}} & & -- & --    \\
M/H  & $-0.01$ & \multicolumn{1}{r|}{ \cite{Sarmento2021}}    & \Mstar\ & $0.47$  & \multicolumn{1}{r|}{ \cite{Kervella2022}}    & & -- & --    \\
Fe/H  & $-0.01$ & \multicolumn{1}{r|}{ \cite{Terrien2015} (b)}  & \Mstar\ & $0.573\pm0.045$  & \multicolumn{1}{r|}{ \cite{Muirhead2018}}    & & -- & --    \\
Fe/H  & $+0.08$ & \multicolumn{1}{r|}{ \cite{Terrien2015} (c)}  & & -- & \multicolumn{1}{r|}{--}    & & -- & --    \\
Fe/H  & $+0.22$ & \multicolumn{1}{r|}{ \cite{Terrien2015} (d)}  & & -- & \multicolumn{1}{r|}{--}    & & -- & --    \\ \hline
\multicolumn{9}{c}{\textbf{Values Derived In This Work}} \\ \hline
\textbf{Param} & \textbf{Value}   & \multicolumn{1}{r|}{\textbf{Reference}} & \textbf{Param} & \textbf{Value}   & \multicolumn{1}{r|}{\textbf{Reference}} & \textbf{Param} & \textbf{Value} & \multicolumn{1}{r}{\textbf{Reference}} \\ 
\hline
Fe/H  & $+0.04$ & \multicolumn{1}{r|}{this work (e)}   & \Mstar\ & 0.454 $\pm$ 0.018 & \multicolumn{1}{r|}{this work} & \Rstar\ & $0.4555\pm0.0128$   & this work \\
Fe/H  & $+0.34$ & \multicolumn{1}{r|}{this work (f)}   & & -- & \multicolumn{1}{r|}{--}    & & -- & --    \\
Fe/H  & $+0.19$ & \multicolumn{1}{r|}{this work (g)}   & & -- & \multicolumn{1}{r|}{--}    & & -- & --    \\
Fe/H  & $+0.09$ & \multicolumn{1}{r|}{this work (h)}   & & -- & \multicolumn{1}{r|}{--}    & & -- & --    \\ \hline
\multicolumn{9}{c}{\textbf{Adopted Values}} \\ \hline
Fe/H  & $+0.3 \pm 0.1$   & \multicolumn{1}{r|}{this work} & Mstar & 0.454 $\pm$ 0.018 & \multicolumn{1}{r|}{this work} & \Rstar\ & 0.4555 $\pm$ 0.0128 & this work \\ 
\hline\hline
\end{tabular}
\tablecomments{
(a) Fe/H measurements from two different LAMOST spectra;
(b) $H$-band;
(c) $J$-band;
(d) $K_s$-band;
(e) this work, photometric metallicity calculated using \citet{Bonfils2005} calibration; 
(f) this work, photometric metallicity using \citet{Johnson2009} calibration; 
(g) this work, photometric metallicity using \citet{Schlaufman2010} calibration; 
(h) this work, photometric metallicity using \citet{Neves2013} calibration.}
\end{table*}

After the pioneering attempts by \citet{Bonfils2005} to estimate photometric metallicities for M dwarfs using $VK_{s}$ color-absolute magnitude relations, multiple studies have attempted improve their accuracy \citep[see, e.g.,][]{Johnson2009, Schlaufman2010, Neves2013}. 
All of the calibrations connect M dwarf companions to FGK-type primary stars with measured metallicities. We list the photometric metallicity estimates using these four calibrations in Table \ref{tab:stellar_params} for reference. 

Using the calibration of \citet{Johnson2009}, and adopting the values $V$ = 13.362, $K_{s}$ = 8.758, and $M_{Ks}$ = 5.882 (Table \ref{tab:star}), we find that TOI-1685 is 0.70 mag above their M dwarf main sequence in ($V-K_{s}$) vs. $M_{Ks}$ space, and estimate a photometric metallicity of [Fe/H] = +0.34. Using the calibration of \citet{Schlaufman2010}, we find that the star's colors are significantly red for its absolute magnitude ($\Delta$($V-K_{s}$) = 0.45), consistent with a photometric metallicity of [Fe/H] = +0.19. We note that both the \citet{Schlaufman2010} and \citet{Neves2013} calibrations adopt a mean metallicity for field M dwarfs which is somewhat low ([Fe/H] = $-0.17$).

Using the recent PASTEL compiled catalog of metallicities from \citet{Soubiran2022}, we find that the local G and K dwarfs in SIMBAD with distances of $d$ $<$ 25\,pc ($\varpi$ $>$ 40\,mas) and dwarf-like surface gravities (4 $<$ log\,$g$ $<$ 5) are consistent with having distributions of metallicities [Fe/H] with means, standard errors, and standard deviations  of ([Fe/H] = $-0.06\pm0.01$, $\sigma$=0.21 dex; $N$(G dwarf)=216) and ([Fe/H] = $-0.05\pm0.01$, $\sigma$=0.23 dex; $N$(K dwarf)=282), respectively.

These distributions are immune to outliers as they are the averages of three estimates of the mean $\mu$ (median, Chauvenet-clipped mean, and probit mean), two estimates of the standard error $\sigma_{\mu}$ (error of the true median and standard error of the Chauvenet-clipped mean), and two estimates of the standard deviation $\sigma$ (68\% confidence intervals, and probit estimate of standard deviation)\footnote{For detailed discussions of these statistical estimators, see \citet{Lutz1980}, \citet{Bevington1992}, \citet{Gott2001}.}. Only a handful of metal-poor stars were clipped from these distributions following Chauvenet's criterion \citep{Bevington1992}: three G dwarfs with metallicities between -0.88 and -0.83 and six K dwarfs with metallicities between -1.76 and -1.04, which are likely halo stars. Hence, the distribution of metallicities [Fe/H] for the local G and K dwarfs within $d$ $<$ 25\,pc are nearly identical. 

K dwarfs are similar to the M dwarfs in that we don't expect them to have evolved off the main sequence during the Galaxy's lifetime. The main sequence lifetime for a $\sim$0.86\,\Msun\, K0V star is $\sim$20 Gyr, i.e. older than the universe, and the lifetimes for lower mass stars are even longer. Therefore we expect the local main sequence K and M dwarfs to have similar metallicity distributions \citep[e.g.][]{Johnson2009}. 

If we reevaluate the \citet{Schlaufman2010} calibration using the updated PASTEL adopted mean field dwarf metallicity of [Fe/H] = -0.05 for K dwarfs, one would derive a new photometric metallicity estimate for TOI-1685 of [Fe/H] = $+0.31$. This is almost identical to the estimate that we derived using the \citet{Johnson2009} calibration, and just slightly higher than the recent value published in H21 ([Fe/H] = $+0.27\pm0.12$ dex) which analyzed IRD spectra using measurements of atomic lines and comparing them to synthetic M dwarf spectra.

As a check on the plausibility of such a high metallicity for TOI-1685, we re-examined the trend for spectroscopic metallicities for the metal-rich primaries of stars with M dwarf companions in Table 1 of \citet{Johnson2009}. Their sub-sample of metal-rich M dwarfs includes the stars HD 46375B, HD 38529B, HD 18143C, 55 Cnc B, HD 190360B and Proxima Centauri, which had $\Delta$$M_{Ks}$ offsets above the main sequence of 0.48, 0.67, 0.68, 0.67, 0.43, 0.36 mags, respectively. 

Querying the recent PASTEL compilation of [Fe/H] values from \citet{Soubiran2022}, we find that the metallicities of the primary FGK stars for these M dwarfs had [Fe/H] values of $0.24\pm0.01$, $0.34\pm0.01$, $0.18\pm0.05$, $0.32\pm0.02$, $0.22\pm0.01$ and $0.22$ (adopting the mean for $\alpha$ Cen A and B), respectively. Indeed, after clipping the pair for HD 18143C with the largest metallicity uncertainty ($\Delta$$M_{Ks}$=0.68, [Fe/H]=$0.18\pm0.05$), the correlation between $\Delta$$M_{Ks}$ and [Fe/H] becomes remarkably tight ($\sigma_{[Fe/H]}$ = 0.016 dex) for the remaining 5 stars:

\begin{equation}
[Fe/H] = 0.062 + 0.395\Delta M_{Ks}
\end{equation}

in the limited range 0.36 $<$ $\Delta$$M_{Ks}$ $<$ 0.67 mag.
Using the \citet{Johnson2009} $V-K_{s}$ vs. $M_{Ks}$ calibration, TOI-1685 has $\Delta$$M_{Ks}$ = 0.70 mag, which is slightly more luminous than even this limited sample of very metal-rich stars examined in the 2009 paper.
Extrapolating this trend to the negligibly more luminous $\Delta$$M_{Ks}$ of 0.70 mag would predict a metallicity of [Fe/H] = $+0.34\pm0.02$, i.e. comparable to the other extremely metal-rich M dwarfs HD 38529 B (M2.5V, $V-K_{s}$ = 4.55, $M_{Ks}$ = 5.66, $\Delta$$M_{Ks}$ = 0.67, [Fe/H] = $+0.34\pm0.01$) and 55 Cnc B (M4V, $V-K_{s}$ = 5.48, $M_{Ks}$ = 7.21, $\Delta$$M_{Ks}$ = 0.67, [Fe/H] = $+0.32\pm0.02$). 

We discount other measurements which measure values closer to solar as the color-magnitude position of the star is extremely well-measured and none of the explanations for the star's over luminosity ($\sim$0.7 mag) above the color-magnitude locus for field M dwarfs makes sense with lower metallicities.

Taking into account our photometric metallicity estimates and the value from H21, we adopt a final metallicity value [Fe/H] = $+0.3\pm0.1$.\\

\subsubsection{Stellar Mass \label{sec:mass}}

We can now estimate the star's mass via the \citet{Mann2019} absolute magnitude-mass-metallicity calibration using the star's adopted parameters ($K_s$ = $8.758\pm0.020$, $\varpi$ = $26.5893\pm0.0192$ mas, [Fe/H] = $+0.3\pm0.1$, $A_V$ = 0) which results in a final mass estimate of \Mstar = $0.454\pm0.018$ \Msun\ (3.9\% uncertainty). This estimate is 1.3\%\ smaller than the stellar mass in H21, and 9.0\%\ smaller than the stellar mass in B21 (Table \ref{tab:stellar_params}). The agreement between our value and that from H21 is unsurprising as both rely on the \citet{Mann2019} calibrations and the final H21 stellar metallicity estimate was also somewhat metal-rich ($+0.14\pm0.12$), whereas the value from B21 was calculated using an estimated radius and a radius-mass calibration from \citet{Schweitzer2019} based on eclipsing binary stars.

\subsubsection{Stellar Radius \label{sec:stellar_radius}}

Previously published radius estimates for TOI-1685 are listed in Table \ref{tab:stellar_params}. We estimate the radius of TOI-1685 using the empirical absolute magnitude-metallicity-radius calibration of \citet{Mann2015}. Adopting $M_{Ks}$ = $5.882\pm0.020$ and [Fe/H] = $+0.3\pm0.1$, the \citet{Mann2015} calibration estimates a radius of \Rstar\, = $0.4555\pm0.0128$ \Rsun. We estimate the radius uncertainty (2.8\%) using the quadrature sum of 0.8\%\, from the uncertainties in the absolute magnitude and metallicity, and 2.7\%\, from the rms scatter in the \citet{Mann2015} calibration. This estimate is 0.8\%\ smaller than the stellar radius in H21, and 8.0\%\ smaller than the stellar radius in B21.

\subsubsection{Bolometric Flux, Luminosity, and Effective Temperature \label{sec:lum}}

We estimate the stellar bolometric flux and luminosity through analysis of the spectral energy distributions using two independent codes: \VOSA\footnote{http://svo2.cab.inta-csic.es/theory/vosa/} \citep[Virtual Observatory SED Analyzer;][]{Bayo2008} and \ARIADNE\footnote{https://github.com/jvines/astroARIADNE} \citep[the spectrAl eneRgy dIstribution bAyesian moDel averagiNg fittEr;][]{Vines2022}. We then combine the estimated luminosity with the empirical radius estimate from \S\ref{sec:stellar_radius} to calculate an independent effective temperature.

The \VOSA\ online tool queries databases of photometry, and fits synthetic stellar spectra of varying parameters (e.g., \teff, \logg, [Fe/H]) from different published spectral libraries, while accounting for user defined distance and reddening \citep{Bayo2008}.  
The \VOSA\, query for TOI-1685 yielded 111 published photometric measurements within 5\arcsec\, of the J2000 position of the star, between 0.36 and 22$\mu$. 
We initially omit from further analysis several magnitudes which were very discrepant at the $>$1\,mag level: $z$ and $y$ from Pan-STARRS PS1 and $J$, $H$, and $K$ (2 $K$ mags) from UKIDSS. 
The vast majority of the remaining \VOSA\, magnitudes are ``OAJ" and ``JPAS" values - magnitudes synthesized from spectrophotometrically calibrated \Gaia\, DR3 Bp/Rp spectra \citep{Montegriffo2023} on the photometric system of the OAJ (Observatorio Astrof\'isico de Javalambre) surveys:  
J-PAS \citep[Javalambre Physics of the Accelerating Universe Astrophysical Survey;][]{Benitez2014}\footnote{https://www.j-pas.org/survey}
and J-PLUS \citep[Javalambre-Photometric Local Universe Survey;][]{Cenarro2019}\footnote{https://www.j-plus.es/}. 
Additional data includes measured photometry from Gaia DR3 \citep{GaiaDR3}, APASS \citep{Henden2016}, 2MASS \citep{Cutri2003}, WISE \citep{Cutri2012}, and NEOWISE \citep{Mainzer2011}. 
Given the dense coverage of the OAJ and JPAS photometry based on Gaia DR3 spectrophotometry, we omit several of the synthetic broadband magnitudes which duplicate coverage (e.g. 
Johnson $BVRI$, SDSS $griz$, HST/ACS/WFC F606W and F814W, and Pan-STARRS $y$). 
Despite it defining the blue end of the photometric coverage, we also omitted the synthetic SDSS $u$ photometry from Gaia DR3 from the analysis, as the synthetic photometry for wavelengths shorter than 400\,nm may contain systematic errors \citep{Montegriffo2023}.

\begin{figure}
    \centering
    \includegraphics[width=.5\textwidth]
    {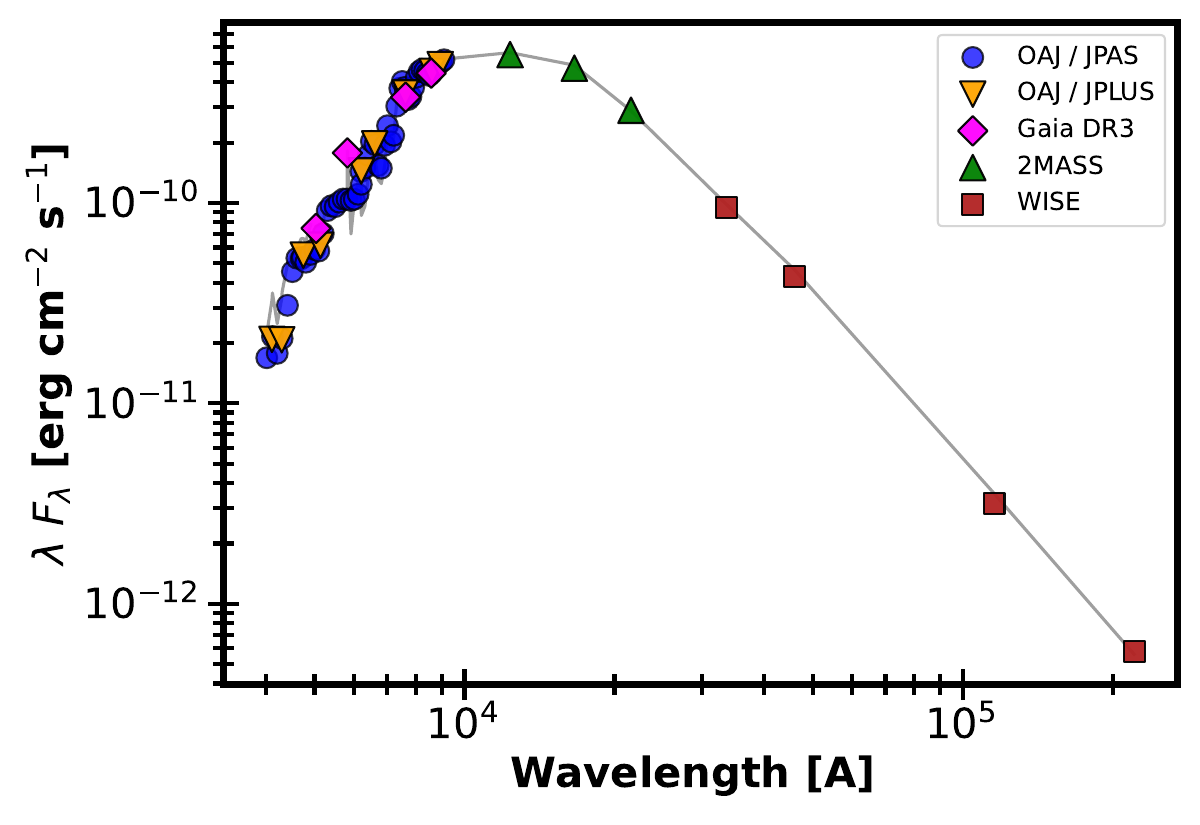}
    \caption{Best fit SED for TOI-1685 from VOSA. Colored points are photometric magnitudes from the Gaia, 2Mass, and WISE surveys, and magnitudes synthesized from spectrophotometrically calibrated \Gaia\, DR3 Bp/Rp spectra into the photometric system of the OAJ JPAS and JPLUS surveys. Grey line depicts the model flux measurements.}
    \label{fig:SED}
\end{figure}

For fitting the SEDs, we adopted zero extinction ($A_V = 0$), a reasonable assumption as at $\sim$38\,pc TOI-1685 is within the Local Bubble \citep[e.g.][]{Reis2011} and modern 3D reddening maps like STILISM \citep{Lallement2018} predict negligible reddening (E(B-V) = $0.001\pm0.014$ mag) at $d$ = 35-40\,pc.  
We used VOSA to fit multiple grids of synthetic stellar spectral models (e.g. 
BT-Settl, BT-Settl/AGSS2009, BT-Settl/CIFIST, BT-Settl/GNS93, BT-Cond, BT-Dusty, BT-NextGen/AGSS2009, BT-NextGen/GNS93) over a a range of stellar temperature and gravity appropriate for main sequence M dwarf stars (2300\,K $<$ \teff\, $<$ 3900\,K; 4.5 $<$ \logg\, $<$ 5.0) and sample over the full range of metallicities covered by the models (although some only used solar).  
The best fit model to the SED of TOI-1685 was a BT-Settl model \citep{Allard2013} with CIFIST2011 solar abundances \citep{Caffau2011} that resulted in \Teff\,=\,$3400\pm50$\,K, \logg\,=\,$5.00\pm0.25$, [M/H]\,=\,0.0 (constrained), with bolometric flux $f_{bol}$ = ($6.90408\pm0.00048$) $\times$ 10$^{-10}$ erg\,s$^{-1}$\,cm$^{-2}$, or $m_{bol}$ = $11.404750\pm0.000075$ on the IAU 2015 bolometric flux/magnitude scale \citep{Mamajek2015} (Figure \ref{fig:SED}). 
This resulted in a stellar bolometric luminosity of $L$ = 0.030524 $L_{\odot}$ ($\pm$0.15\%) or \logL\, = $-1.51537\pm0.00065$ dex. 
The fit used 87 of the 90 data points, and the relative uncertainty in the flux is fantastically tiny (0.0049\%), much smaller than the absolute uncertainties achievable in spectrophotometry. 

The best calibrated spectrophotometric standard stars are absolutely flux
calibrated in the visible and near-IR to $\sim$1\% \citep[e.g.,][]{Bohlin2019,Rieke2022}, and photometric passband zero point fluxes for the classic visible and near-IR bands \citep[e.g., Johnson, 2MASS, etc.;][]{MannvonBraun2015}, Gaia photometry \citep[][]{Pancino2021}, 
and Gaia synthetic photometry \citep[][]{Pancino2022}
all appear to be accurate at the $\sim$1\% level.
Considering that the absolute flux calibration of the SED fit includes uncertainties from all of these sources, we conservatively adopt an \fbol\, uncertainty of 2\%, or $f_{bol}$ = ($6.904\pm0.138$) $\times$ 10$^{-10}$ erg\,s$^{-1}$\,cm$^{-2}$ and $m_{bol}$ =
$11.405\pm0.022$ on the IAU 2015 bolometric flux/magnitude scale \citep{Mamajek2015}. 
This results in a stellar bolometric luminosity of $L$ = $0.03052\pm0.00061$ $L_{\odot}$ or \logL\, = $-1.51537\pm0.00869$ dex. 

\texttt{ARIADNE} uses Bayesian Model Averaging to include information from a variety of stellar models (e.g. BT-Settl, BT-Cond, BT-NextGen, and Phoenix) when determining the best-fit stellar \teff\, \logg\, [Fe/H], and radius. The package's built in search function returns a smaller set of photometry than VOSA, consisting of the Gaia B$_{\rm{P}}$, G, and R$_{\rm{P}}$ bands, 2MASS J, H, and K$_{\rm{s}}$ bands, and the WISE W1 and W2 bands.
The Bayesian fit of the SEDs to the photometry, with the extinction again fixed to zero, is: \
Teff\,=\,$3431^{+83}_{-42}$\,K, 
\logg\,=\,$4.83^{+0.043}_{-0.039}$ dex, 
[Fe/H]\,=\,$0.26^{+0.16}_{-0.19}$, and \Rstar =\,$0.4838^{+0.0173}_{-0.0185}$ \Rsun. 
The corresponding best-fit luminosity is $0.02922^{+0.0034}_{-0.0026}$ \Lsun, which is 4.3\%\, (2.2-$\sigma$) lower than the VOSA estimate above.
The BT-Settl model is preferred over the Phoenix model in a 69\% to 31\% probability ratio. 

Through fitting a NextGen model to 10 broadband photometry data points, H21 estimated \fbol = ($6.65\pm0.15$) $\times$ $10^{-10}$ erg\,s$^{-1}$\,cm$^{-2}$ (\mbol\, = $11.445\pm0.024$ mag on IAU scale) and \Lbol\, = $0.0271^{+0.0028}_{-0.0026}$ \Lsun (their Table 1).
The luminosity for TOI-1685 quoted by H21 in their Table 1 is notably 11.2\%\, lower than what we estimate, and at 3.7-$\sigma$ relative to the uncertainties we and they quote in \Lbol. 
Oddly, if one combines the Gaia DR3 distance (37.609\,pc) and \fbol\, in H21's Table 1 through $L$ = 4$\pi$$d^2$\fbol, one would estimate a stellar luminosity of \Lbol\, = 0.0294 \Lsun, remarkably 8.5\% higher than the luminosity quoted in their Table 1, and only 3.7\%\, lower than our estimated luminosity from the VOSA SED fit.
We conclude that the luminosity for TOI-1685 in H21 is likely underestimated by $\sim$11\%, and that the actual agreement in \fbol\, and \Lbol\, estimates between our VOSA SED analysis and the NextGen SED fit from H21 is actually at the $\sim$3.7$\pm$3.0\% (1.2-$\sigma$ level), i.e. consistent. 

Combining our new luminosity estimate ($L$ = $0.03052\pm0.00061$ $L_{\odot}$) with the empirical radius estimate derived in \S \ref{sec:stellar_radius} (\Rstar\, = $0.4555\pm0.0128$ \Rsun), we derive an independent effective temperature which we adopt for the rest of the paper's analyses and discussions: \Teff\, = $3575\pm53$\,K.
This value is within 150\,K of the value derived from the ARIADNE SED analysis. Table \ref{tab:teffs} summarizes previously published \teff\, estimates for TOI-1685 and the values span nearly 20\%. The median of these previously published estimates is \Teff\, = 3502\,K, just 1.4-$\sigma$ lower than our new VOSA-based value. Further, our new estimate is within 2-$\sigma$ of several published estimates \citep{Terrien2015, Muirhead2018, Bai2018, Birky2020, Sebastian2021, Hirano2021, Sprague2022, Queiroz2020, Jonsson2020, Yu2023} based on multiple methods (analysis of visible or near-IR spectra, SED fits, colors-\Teff\, relations).
%

\begin{table*}[htbp]
\caption{Effective Temperature Estimates for TOI-1685}\label{tab:teffs}
\scriptsize
\centering
\begin{tabular}{lrlr}
\hline \hline
\multicolumn{1}{c}{\textbf{Value}} & \multicolumn{1}{c|}{\textbf{Reference}} & \multicolumn{1}{c}{\textbf{Value}} & \multicolumn{1}{c}{\textbf{Reference}} \\ 
\hline
\multicolumn{4}{c}{\textbf{Published Literature Values}} \\ 
\hline
$3434\pm51$\,K & \multicolumn{1}{r|}{\citet{Bluhm2021}} &  $3660\pm9$\,K &  \citet{Olney2020}  \\
$3428\pm97$\,K & \multicolumn{1}{r|}{\citet{Hirano2021}$^a$} & $3450$\,K & \citet{Morrell2019} \\
$3475\pm75$\,K & \multicolumn{1}{r|}{\citet{Hirano2021}$^b$} & $3457\pm157$\,K & \citet{Stassun2019} \\
$3429\pm114$\,K & \multicolumn{1}{r|}{\citet{Hardegree-Ullman2023}} & $3463^{+24}_{-29}$\,K & \citet{Anders2019} \\
$3612\pm98$\,K & \multicolumn{1}{r|}{\citet{Yu2023}} &$3986\pm33$\,K & \citet{Xiang2019}$^h$ \\
$3594\pm10$\,K & \multicolumn{1}{r|}{\citet{Sprague2022}} & $4165\pm47$\,K & \citet{Xiang2019}$^h$ \\
$3629\pm105$\,K & \multicolumn{1}{r|}{\citet{Sebastian2021}} & $3490\pm63$\,K & \citet{Muirhead2018} \\
$3504$\,K & \multicolumn{1}{r|}{\citet{Birky2020}$^c$} &$3624$\,K & \citet{Bai2018} \\
$3615\pm157$\,K & \multicolumn{1}{r|}{\citet{Zhang2020}$^d$} & $3473$\,K & \citet{Terrien2015}$^i$ \\
$3579\pm138$\,K & \multicolumn{1}{r|}{\citet{Zhang2020}$^e$} & $3481$\,K & \citet{Terrien2015}$^j$\\
$3596^{+9}_{-11}$\,K & \multicolumn{1}{r|}{\citet{Queiroz2020}} &$3499$\,K & \citet{Terrien2015}$^k$ \\ 
$3600$\,K & \multicolumn{1}{r|}{\citet{Jonsson2020}$^f$} & $3555$\,K & \citet{Terrien2015}$^l$ \\ 
$3688\pm60$\,K & \multicolumn{1}{r|}{\citet{Jonsson2020}$^g$}  & $3888^{+202}_{-417}$\,K & Gaia DR2 \\
\hline
\multicolumn{4}{c}{\textbf{Values Derived In This Work}} \\ 
\hline
\textbf{3575 $\pm$ 53 K} & \multicolumn{1}{r|}{ \textbf{this work$^{m}$, adopted}} &  $3431^{+83}_{-42}$ K &  this work$^{n}$  \\ 
\hline \hline
\end{tabular}
\tablecomments{For papers that contain more than one method for estimating \Teff\, we use the following notes to clarify which value is reported in this table : 
(a)	\citet{Hirano2021} IRD spectrum analysis;
(b)	\citet{Hirano2021} NextGen SED analysis;
(c)	\citet{Birky2020} ASPCAP pipeline;
(d)	\citet{Zhang2020} 1st APOGEE/SLAM fit, however it gives unrealistically low gravity (log$g$ $\sim$ 1) and metallicity ([M/H] $\sim$ -0.9);
(e)	\citet{Zhang2020} 2nd APOGEE/SLAM fit, however it gives unrealistically low gravity (log$g$ $\sim$ 1.2) and metallicity ([M/H] $\sim$ -0.5);
(f)	 \citet{Jonsson2020} APOGEE-2;
(g)	\citet{Jonsson2020} APOGEE-2/ASPCAP;
(h)	 \citet{Xiang2019} estimates from two independent spectra, however fits give unrealistic gravity and metallicity (log$g$ $\simeq$ 4 and [Fe/H] $\simeq$ -1.4);
(i)	\citet{Terrien2015} K-index;
(j)	\citet{Terrien2015} \citet{Newton2015} calibration;
(k)	\citet{Terrien2015} H-index;
(l)	\citet{Terrien2015} J-index;
(m)	derived from VOSA luminosity estimate combined with updated empirical radius estimate from \citealt{Mann2015};
(n)	\texttt{ARIADNE} SED fit results.
}
\end{table*}

\subsubsection{Stellar Rotation and Age}
\label{sec:rotage}

Our best-fit radial velocity analysis of TOI-1685 yielded a quasi-periodic signal with a period posterior distribution of $18.2 \pm 0.5$ days, similar to the 19 day signal observed in B21.  We interpret this to represent the rotation of the star.  

Main sequence dwarf stars spin down by losing angular momentum through a magnetized wind, and under certain conditions a rotation period can be used to estimate the star's gyrchronologic age. Namely if it has spun down sufficiently such that the Rossby number (ratio of rotation period to convective turnover time) is larger than a critical value, its magnetic activity is below the ``saturation" level, and a rotation-dependent braking law erases differences in initial spin rates and causes convergence to a single temperature-dependent rotation sequence.  Models are not advanced enough to produce robust braking laws applicable to different stellar types, and empirical calibration using observed rotation sequences at appropriate ages is needed. Germane to TOI-1685, \citet{Dungee2022} established a rotation sequence for early type M dwarfs in the 4 Gyr-old cluster M67.  Combining this with younger sequences \citep[e.g.,][]{Curtis2020}, \citet{Gaidos2023} assigned ages with formal uncertainties to the late K and early M-type host stars of many known exoplanets.  The primary sources of error are uncertainties in rotation period and \teff, residual star-to-star differences in spin, and the effect of metallicity (which affects the moment of inertia and also perhaps the braking law)\footnote{The metallicity of M67 and the young clusters used to construct the gyrochronology in \citet{Gaidos2023} are all close to solar}.  Using the procedures described in \citet{Gaidos2023}, we estimate an age of $1.0 \pm 0.3$ Gyr for TOI-1685, with a model-based correction for [Fe/H]=+0.3 of 0.3 Gyr \citep[i.e., the star's actual age could be $\sim$1.3 Gyr, see Fig. 5 in ][]{Gaidos2023}.

\subsection{Transit Photometry Analyses \label{sec:transit_phot}}

To jointly model all of the \TESS\ data (see Sec. \ref{sec:TESS}) and various ground-based photometric light curves (see Sec. \ref{sec:ground}), we used the sum of components modelling the transit of TOI-1685 b, quasi-periodic modulations arising due to stellar activity, and systematic noise originating in each instrument. To achieve this we used \texttt{juliet} \citep{Espinoza2019}, a python package that facilities the simultaneous modelling of transit data acquired in multiple bandpasses.

For the transit, we used a \citet{MandelAgol:2002} transit model generated using the \texttt{batman} package \citep{Kreidberg2015}, and parameterized using the period $P$, transit time $t_\mathrm{0}$, radius ratio $R_\mathrm{p}/R_\star$, impact parameter $b$, and stellar density $\rho_\star$, with eccentricity $e$ fixed to 0. We used the quadratic law to model limb darkening, following the $q$ parameterization of \citet{Kipping2013}. Different light curves acquired using a common bandpass (see ``Band'' column in Table \ref{tab:ground}) shared a single pair of limb darkening coefficients.

For each of the ground-based light curves, we included auxiliary observing parameters such as airmass and detector $x-y$ offset in the model to decorrelate systematic noise, with the parameters used for each light curve shown by the ``Detrending'' column in Table \ref{tab:ground}.

Quasi-periodic photometric variability with a timescale of a few days was apparent by eye in both \TESS\ sectors, in line with expectations for stellar variability of early M-type stars. However both the period and amplitude of the variability change significantly in the intervening two years between the two \TESS\ sectors. The Lomb-Scargle periodogram indicated a 10 d periodicity that was present in both sectors, along with a 4.5 d periodicity that was only present in sector 59. For this reason we used two different Gaussian Processes \citep[GPs;][]{Gibson2014} using simple harmonic oscillator kernel generated using the \texttt{celerite} package \citep{ForemanMackey2017} with quality factor $Q$ -- which dictates the coherency of the oscillations -- set to $1/\sqrt{2}$. The first GP was designed to fit the 10 d periodicity that was relatively consistent in both sectors, while the second GP fit only the additional 4.5 d periodicity in sector 59. 

There was residual time-correlated noise on timescales longer than the observations in many of the ground-based light curves even after detrending against the auxiliary observing parameters, likely originating from the same stellar variability. We tested fitting GPs with a range of periods covering those that we observed in both sectors of \TESS\ data to the ground-based light curves and found them to be broadly consistent with all of them, mainly due to the limited time baseline of the light curves. As both GPs used to fit the \TESS\ data described the residual noise in the ground-based light curves well, and to avoid adding unnecessary extra degrees of freedom to the baseline models of the ground-based light curves, we extended the 10-day GP described in the paragraph above to jointly model the variability in all of the ground-based light curves as well.

\begin{figure}
    \centering
    \includegraphics{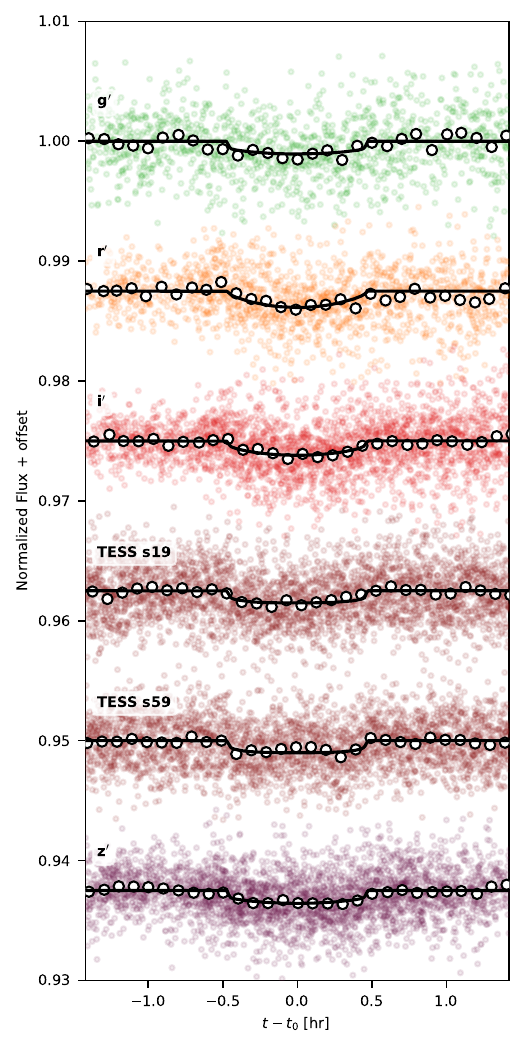}
    \caption{All of the light curves used in the global transit fit of TOI-1685 b, detrended and phase folded by common photometric bandpass. The two \TESS\ sectors are shown separately. The median transit models are shown with black lines, and the light curves in 6 minute bins are shown as white circles, with error bars generally too small to be visible.}
    \label{fig:transits}
\end{figure}

To derive the joint posterior distributions for each free parameter, we used the Dynamic Nested Sampler \citep[see][]{Skilling2004,Skilling2006,Higson2019} from the \texttt{dynesty} package \citep{Speagle2020}. We used the new measurements of the stellar mass and radius presented respectively in Sections \ref{sec:mass} and \ref{sec:stellar_radius} to place a Gaussian prior on $\rho_\star$. As \citet{Patel2022} found a significant discrepancy between theoretically predicted and empirical limb darkening coefficients for M-dwarfs, we selected uniform priors that allowed all physically possible values to be sampled (i.e. with limits at 0 and 1). We placed wide uniform priors on $P$, $t_\mathrm{0}$, $R_\mathrm{p}/R_\star$, $b$ and the linear coefficients of the auxiliary observing parameters, along with wide uniform priors on the logarithm of the GP hyperparameters. The results for a selection of transit and physical parameters are displayed in Table \ref{tab:SysUpdate}, with the other free parameters displayed in Table \ref{tab:noise}. The final best-fit transit depth is $\Delta F$ = $874^{+34}_{-36}$ ppm, which is 14.4\% smaller than the 1 ppt depth measured by H21 and B21 (Figure \ref{fig:transits}).

Combining the updated stellar radius estimate of \Rstar\ = $0.4555 \pm 0.0128$ \Rsun\ with our best fit transit depth produces an updated planet radius measurement of 1.468 $\pm$ 0.05 \Rearth\ for TOI-1685 b.


\begin{table*}
\begin{center}
  \caption{Selection of the updated stellar and system parameters from the stellar characterization described in \S\ref{sec:stellar_params}, the transit photometry fit described in \S\ref{sec:transit_phot} \& the 1D Gaussian Process RV fit described in \S\ref{sec:rv_analysis}. Additional fit results can be found in the Appendix.}  
  \label{tab:SysUpdate}
  \begin{tabular}{lcccc}
  \hline
  \hline
  \noalign{\smallskip}
  Parameter & & Adopted Value & Comparison$^{(a)}$ &  Comparison \\
   & &  & to B21 &  to H21 \\
  \noalign{\smallskip}
  \hline\hline
  \noalign{\smallskip}
  \multicolumn{5}{l}{\emph{\bf Adopted stellar parameters}} \\
   Stellar Metallicity [Fe/H]  & -- & 0.3 $\pm$ 0.1 & $\uparrow$ 143\% & $\uparrow$ 53\% \\
   Stellar Radius [\Rsun]  & -- & 0.4555 $\pm$ 0.0128 & $\downarrow$ 8.0\% & $\downarrow$ 0.8\% \\
   Stellar Mass [\Msun]  & -- & 0.454 $\pm$ 0.018 & $\downarrow$ 9.0\% & $\downarrow$ 1.3\% \\
   Stellar F$_{bol}$ [erg s$^{-1}$ cm$^{-2}$]  & -- & (6.904 $\pm$ 0.138) x 10$^{-10}$ & NR$^{(b)}$ & $\uparrow$ 3.7\% \\
   Stellar Luminosity [\Lsun]  & -- & 0.03052 $\pm$ 0.00061 & $\uparrow$ 0.7\% & $\uparrow$ 3.7 / $\uparrow$11.2\%$^{(c)}$  \\
   Stellar \Teff\ [K]  & -- & 3575 $\pm$ 53 & $\uparrow$ 3.9\% & $\uparrow$ 3.2\% \\
    Stellar Age [Gyr]  & -- & 1.0 $\pm$ 0.3 & NA & NA \\
    & & & & \\
  \hline\hline
  \noalign{\smallskip}
  Parameter & Prior$^{(d)}$ & Final value$^{(e)}$ & Comparison &  Comparison \\
   & &  & to B21 &  to H21 \\
  \hline\hline
  \noalign{\smallskip}
  \multicolumn{5}{l}{\emph{\bf Sampled transit parameters }} \\
  \noalign{\smallskip}
    $\rho_\star$ [g cm$^{-3}$] & $\mathcal{N}[6.77,0.63]$ & $6.55^{+0.21}_{-0.25}$ &  $\uparrow$ 11.5\% & $\downarrow$ 2.3\%  \\
    Orbital period $P_{\mathrm{orb}}$ [days]  & $\mathcal{U}[0.668,0.670]$ & $0.66913923^{+0.00000040}_{-0.00000038}$ & 0\% & 0\% \\
    Transit epoch $T_0$ [BJD$_\mathrm{TDB}-$2450000]  & $\mathcal{U}[2459910.92,2459910.94]$ & $9910.93830^{+0.00037}_{-0.00038}$ & N/A$^{(f)}$ & N/A \\
    $R_\mathrm{p}/R_\star$ & $\mathcal{U}[0.01,0.04]$ & $0.02956^{+0.00057}_{-0.00061}$ & $\downarrow$ 7.2\% & $\uparrow$ 1.6\% \\
    $b$ [R$_\star$] & $\mathcal{U}[0,1]$ & $0.266^{+0.062}_{-0.075}$ & N/A & N/A  \\
  \noalign{\smallskip}
  \multicolumn{5}{l}{\emph{\bf Sampled RV parameters }} \\
  \noalign{\smallskip} 
    Orbital period $P_{\mathrm{orb}}$ [days]  & $\mathcal{N}[0.66913923 , 0.00000040]$ & $ 0.66913924 \pm 0.00000042 $ & 0\% & 0\% \\
    Transit epoch $T_0$ [BJD$_\mathrm{TDB}-$2450000]  & $\mathcal{N}[9910.93830 , 0.00038]$ & $ 9910.93828_{-0.00039}^{+0.00041} $ & N/A & N/A \\  
    Orbital eccentricity, $e$ & $\mathcal{F}[0]$ & 0 & locked to 0 & locked to 0 \\
    Doppler semi-amplitude $K$ [m s$^{-1}$] & $\mathcal{U}[0,50]$ & $ 3.76_{-0.38}^{+0.39} $ & $\downarrow$ 17.3\% & $\downarrow$ 11.7\% \\
     & & & & \\
    \hline
    \multicolumn{5}{l}{\emph{\bf Derived planet parameters}} \\
    \noalign{\smallskip}
    $\Delta F$ [ppm] & -- & $874^{+34}_{-36}$ & $\downarrow$ 14.4\% & $\downarrow$ 14.4\% \\
    $t_\mathrm{14}$ [hr] & -- & $0.918\pm0.015$ & NR & NR \\
    Semi-major axis [AU] & -- & $0.01138^{+0.00035}_{-0.00034}$ & $\downarrow$ 3.7\% & $\downarrow$ 1.6\% \\
    Orbital inclination [deg] & -- & $87.17^{+0.81}_{-0.70}$ & $\uparrow$ 2.8\% & $\uparrow$ 1.8\% \\
    Planet Radius [R$_\oplus$] & -- & $1.468^{+0.050}_{-0.051}$ & $\downarrow$ 15.8\% & $\uparrow$ 0.6\%  \\
    Planet Mass [\Mearth] &  -- & $ 3.03_{-0.32}^{+0.33}$ & $\downarrow$ 24.8\% & $\downarrow$ 13.2\%  \\
    Planet Density [g cm$^{-3}$] & -- & 5.3 $\pm$ 0.8 & $\uparrow$ 20.6\% & $\downarrow$ 15.1\% \\
    Planet Insolation [S$_{\rm{\oplus}}$] & -- & 236 $\pm$ 15 & $\downarrow$ 8.1\% & NR \\
    Planet Equilibrium Temperature [K] & -- & 1062 $\pm$ 27 & N/A & N/A \\
    \hline
\multicolumn{5}{l}{\footnotesize $^a$ an upward arrow in the comparison column indicates that our final value is X\% larger than the B21 or H21 value, while a downward arrow} \\ 
\multicolumn{5}{l}{\footnotesize indicates that our final value is X\% smaller than the B21 or H21 value. } \\
\multicolumn{5}{l}{\footnotesize $^b$ Comparison not possible because value was not reported in the original publication's summary tables or text}\\
\multicolumn{5}{l}{\footnotesize $^c$ As reported in H21's Table 1, the L values differ by 11.2\%, but see \S\ref{sec:lum} of this work for discussion on why the offset is likely only 3.7\%}\\
\multicolumn{5}{l}{\footnotesize $^d$ $\mathcal{F}[a]$ refers to a fixed value $a$, $\mathcal{U}[a,b]$ to an uniform prior between $a$ and $b$, $\mathcal{N}[a,b]$ to a Gaussian prior with mean $a$ and standard deviation $b$.} \\
\multicolumn{5}{l}{\footnotesize $^e$ Inferred parameters and errors are defined as the median and 68.3\% credible interval of the posterior distribution.}\\
\multicolumn{5}{l}{\footnotesize $^f$ Comparison not applicable due to mismatch in units or reporting metric with the original publication's summary tables or text.}\\
  \hline
  \hline
  \end{tabular}
\end{center}
\end{table*}

\subsection{Radial Velocity Analyses \label{sec:rv_analysis}}

To jointly model the combined CARMENES, IRD, MAROON-X Blue, and MAROON-X Red data sets, we used Pyaneti \citep{Barragan2019, Barragan2022}, an open source python package. Pyaneti can perform traditional 1-D Gaussian Process fits, multi-dimensional Gaussian Process fits, and Floating Chunk Offset fits \citep[FCO;][]{Hatzes2011}. 

One current restriction of Pyaneti is that RV data sets must be square matrices, meaning that each instrument must have the same of input columns (date, RV, RV error, activity index, etc). While the CARMENES and MAROON-X data sets both produce estimates of the differential line width (dLW) of the RV spectra and spectral activity indicators derived using the H-$\alpha$ and Calcium IR triplet absorption lines, the IRD data set does not contain any overlapping activity indices. Therefore opting for a multi-dimensional GP model would require removing the IRD data from the fit.

To assess this option we performed a series of two-dimensional GP fits that incorporate just the MAROON-X and CARMENES data sets, first using the RVs and the dLW time series, then using the RVs and the H-$\alpha$ time series, and finally the RVs and the Ca IR triplet time series. We find that the dLW measurements are not informative enough to further constrain the GP, as they exhibit significant scatter around the best fit model.
RV instruments and data reduction pipelines are built to be most sensitive to the centroids of absorption lines (used to measure the star's RV) rather than to asymmetries in the lines' shapes (captured in the dLW). It is thus not surprising that small line asymmetries produced by a relatively inactive star such as TOI-1685 do not produce signals substantial enough to help constrain the GP. 
Similarly, 2D GPs that incorporate either the H-$\alpha$ or Ca NIR triplet measurements do not provide significant improvements over the simpler 1D model where we are able to include all three RV instruments' time series. Periodograms of the combined CARMENES and MAROON-X H-$\alpha$ and Ca IR triplet time series do not exhibit singular, significant, peaks near the rotation period of the star, and so their lack of constraining power when added to the GP model is to be expected. Knowing this, we adopt the simpler 1-D Gaussian Process model which considers the RV time series of all three instruments without any simultaneous stellar variability metrics when determining the lifetime, evolution timescale, period, and amplitude of the Gaussian Process that is fit alongside the Keplerian planet model.

As both of the discovery papers suggest the potential for an additional planet in the system (at periods of P = 2.6 days and P = 9.025 days in H21 and B21, respectively) we test five different planetary system models. Each model includes 1 circular planet initiated at the best-fit period of TOI-1685 b from the joint photometry fit in \S\ref{sec:transit_phot}. The models then have either no additional planet, or one additional planet initiated with the orbital parameters presented for the planet candidate in either B21 and H21, which is either restricted to a circular orbit or allowed to be eccentric. We fix the eccentricity of TOI-1685 b to zero as observational evidence from the Kepler mission shows that small planets with orbital periods shorter than $\sim$5 days have eccentricities consistent with zero, as expected due to tidal circularization \citep[see, e.g,][]{Shabram2016, VanEylen2019}.

We apply a Gaussian prior on planet b's period and transit epoch taken from the updated \TESS\ + Muscat fits above. For the two planet candidates we adopt Gaussian priors on the period and epoch taken from their respective announcement publications. Each model also includes a quasi-periodic GP initiated with a uniform prior on the period from 17 - 20 days to address the star's rotational variability. We also implement one Floating Chunk Offset fit, which is a single planet model that includes only the USP planet without any GP component as the nightly offsets are expected to remove the effects of any stellar variability occurring on timescales longer than $\sim$10 hours \citep[see, e.g.,][]{Dai2017}.

\begin{figure*}[ht!]
    \centering
    \includegraphics[width=.9\textwidth]{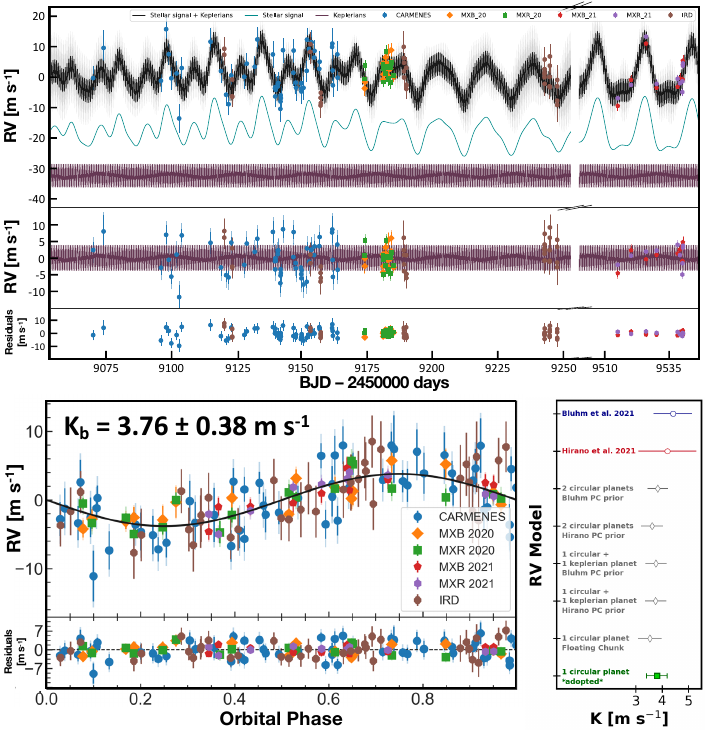}
    \caption{Top: Time series RV data used in our adopted 1 circular planet + 1D quasi-periodic Gaussian Process fit, color coded by instrument. The MAROON-X data sets are broken up between the 2020 and 2021 semesters due to interventions in the spectrograph that introduce a zero-point offset. The purple line is the model of planet b's orbit, the teal line is the GP representing the star's contribution to the RV signal, and the black line is the planet's orbit combined with the GP. The light-shaded areas show the GP model’s one- and two-sigma credible intervals. Bottom Left: RV data phase folded to the best-fit period of planet b. Bottom Right: Comparison of the best-fit semi-amplitudes from the published literature (blue and red points, depicting the results from B21 and H21, respectively) and the different pyaneti models we tested on the combined IRD + CARMENES + MAROON-X data set. Models with two planets always assume a circular orbit for planet b, and then either a circular (`cp') or Keplerian (`kp') orbit for the outer planet. We place Gaussian priors on the outer planet candidate's period and transit epoch using the reported planet candidate parameters from B21 and H21. All models produce consistent results for the semi-amplitude of planet b, but the preferred solution according to comparisons of the Akaike Information Criteria is a single circular planet plus a 1D quasi-periodic GP, which is depicted in green.}
    \label{fig:RV_Summary}
\end{figure*}

All six fits produce good agreement in the best-fit period and semi-amplitude values for planet b (bottom right panel of Figure \ref{fig:RV_Summary}). To decide which model to adopt, we must first select an appropriate metric for comparing their performance and goodness of fit. We note that the commonly cited Bayesian Information Criterion (BIC) is seen to be most useful in selecting the true model for a given system, but this requires the assumption that the true model is included in the set of tested models. Meanwhile the Akaike Information Criterion (AIC) is better at selecting the \textit{best} model out of those included in the test set, even if the \textit{true} model is not included. For more on this distinction see, e.g., \citealt{burnham2002model} and \citealt{Chakrabarti2011}. As it is reasonable to assume that we do not have complete knowledge of all planets nor stellar variability components for any given exoplanet system, we opt use the AIC as our model comparison metric. When comparing AIC values across the six models described above we find that the single, circular planet model combined with a GP to address the stellar activity is best supported by the data, and so we adopt this as our fiducial RV model. 

In our final RV fit, we adopt Gaussian priors on planet b's period and transit epoch from the best-fit transit results in \S\ref{sec:transit_phot} and lock the planet's eccentricity to zero. We adopt a uniform prior of 17 - 20 days on the GP period hyperparameter ($P_{GP}$) based on the rotation period reported in B21 and the rotational modulation seen in our combined RV data set, and another uniform prior of 1 - 150 days on the spot lifetime hyperparameter ($\lambda_e$) as this corresponds to our 2020 observing baseline and we do not expect the two observing seasons to be strongly correlated. This produces a best-fit semi-amplitude of $K_{\mathrm{b}} = 3.76^{+0.39}_{-0.38}$ \ms\ (Table \ref{tab:SysUpdate}) which is 11.7\% and 17.3\% smaller than the best-fit K values from H21 and B21, respectively. The full set of RV fit results is presented in the Appendix in Table \ref{tab:pyaneti}.

When combined with the updated stellar mass estimate from \S\ref{sec:stellar_params}, \Mstar\ = $0.454 \pm 0.018$ \Msun\ , and the orbital inclination angle derived from the transit photometry fit, $i$ = 87.17$^{+0.81}_{-0.70}$ degrees, our best-fit RV semi-amplitude produces an updated planet mass measurement of $M_{\mathrm{b}}$ = 3.03$^{+0.33}_{-0.32}$ \Mearth\ for TOI-1685 b.

\section{Discussion} \label{sec:discussion}

\subsection{Updates to Planet Parameters \label{sec:update}}

\begin{figure*}[ht!]
    \centering
    \includegraphics[width=.95\textwidth]{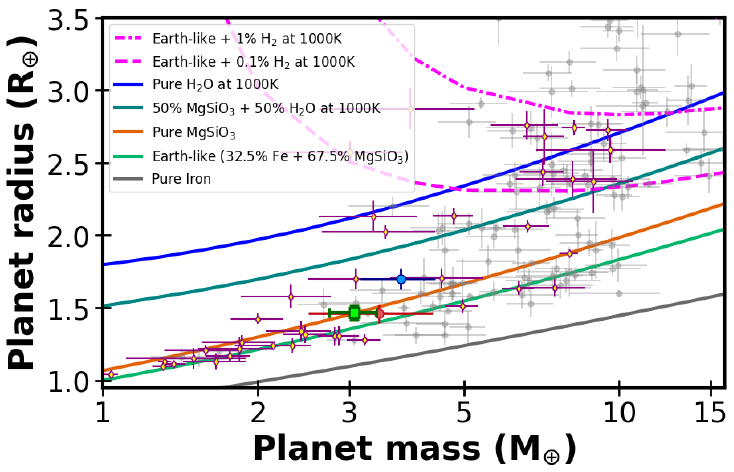}
    \caption{Small planet mass vs radius diagram, using data taken from NASA's Exoplanet Archive on October 17, 2023. Grey points with error bars show confirmed planets with mass and radius measurements better than 20\%,  solid lines represent two-layer models as given by \citet{Zeng2016}. Large colored circles are the previous characterizations of TOI-1685 b from B21 in blue and H21 in red, while the green square depicts the best-fit mass and radius for the planet derived here. Purple diamonds show planets that have been selected for observation with \JWST\ in Cycle 1 or 2.}
    \label{fig:MassRadiusDiagram}
\end{figure*}

Combining the best-fit planet radius from \S\ref{sec:transit_phot}, R$_{\mathrm{b}}$ = 1.468$^{+0.050}_{-0.051}$ R$_{\oplus}$, with the best-fit planet mass from \S\ref{sec:rv_analysis}, $M_{\mathrm{b}}$ = 3.03$^{+0.33}_{-0.32}$ \Mearth, produces a best-fit planet density of 5.3 $\pm$ 0.8 g cm$^{-3}$. The similarity to the Earth's bulk density of 5.51 g cm$^{-3}$ suggests that TOI-1685 b is also a rocky planet with minimal volume contributions from any Hydrogen or Helium atmosphere components (Figure \ref{fig:MassRadiusDiagram}).

The RV semi-amplitude measured from our combined CARMENES, IRD, and MAROON-X data set is smaller than the values reported in either B21 or H21 (Table \ref{tab:SysUpdate}). While our new value is within 1-sigma of both previously published results, the 10-20\% decrease in measured RV semi-amplitude as more data was added highlights the importance of continued observing, especially in systems with low mass planets that induce small RV signals. Planets' whose RV semi-amplitudes are measured to be slightly too high are more likely to get published than planets whose RV semi-amplitudes are measured to be slightly too low when striving to meet 3- or 5-$\sigma$ detection significance criteria \citep[see, e.g.,][]{Burt2018, Montet2018}, which can in turn bias the small planet end of the exoplanet mass-radius distribution.

Our measured R$_{\mathrm{b}}$ is in very close agreement with H21, and disagrees with B21 and L22 at the 2.7-$\sigma$ level. This comes in large part from the disagreement in R$_\star$ values derived by B21 (which was also adopted by L22) and H21. The R$_\star$ that we independently derived in \S\ref{sec:stellar_radius} is within 0.3-$\sigma$ of the best-fit stellar radius from H21, but it is 1.9-$\sigma$ smaller than the \Rstar\ value from B21. 
Our smaller stellar radius contributes to a smaller planet radius as derived in \S\ref{sec:transit_phot} and a higher \Rhopl\ when combining the best-fit planet radius and mass. Another contribution to the planet radius disagreements could be the fact that neither of the two nights of MuSCAT2 data that were used in the B21 transit fit passed our screening process (described in \S\ref{sec:ground}) to be included in our joint transit fit, and both of these appear to favour a much larger planet radius. 

The updated stellar radius and effective temperature also flow down to impact TOI-1685 b's expected insolation and equilibrium temperature values. The planet's updated stellar insolation, S = L/a$^{2}$, is 236 $\pm$ 15 S$_{\oplus}$. Assuming a Bond Albedo of A$_{\rm{B}}$ = 0.1, as suggested by both laboratory astrophysics studies \citep[see, e.g.,][]{Essack2020} and recent \JWST\ observations of the USP super-Earth GJ 367 b \citep{Zhang2024}, TOI-1685 b's equilibrium temperature is 1062 $\pm$ 27 K. Given the predominantly MgSiO$_{\rm{3}}$ composition suggested by the planet's position on the mass-radius diagram (Figure \ref{fig:MassRadiusDiagram}), the internal composition analysis carried out in \S\ref{sec:internal_comp}, and the fact that silicate rocks begin melting at 850 K \citep{Lutgens2014} TOI-1685 b is likely to be a lava world with a molten dayside surface.

\subsection{Interpretation of previous planet candidates}

In situ formation for USP planets like TOI-1685 b is deemed unlikely, as their P\textless 1 day orbits lie interior to the dust sublimation radius of typical protoplanetary disks \citep{Millholland2020}. Current formation models for these objects generally invoke dynamical interactions with another planet in the system, which drives the USP from a longer period formation location to its current orbital position \citep[see, e.g., ][]{Pu2019, Petrovich2019, Millholland2020}. An alternative theory puts forth that USPs are the remnant cores of of Hot Jupiter or Hot Saturn progenitors \citep{Jackson2013, Valsecchi2014, Jackson2016}, but the lack of correlation between USP formation and increased stellar metallicity has resulted in this explanation generally being discarded by the community \citep{Winn2017}. It is therefore reasonable to expect that the TOI-1685 system hosts at least one additional planet that helped to shape the current day orbit of TOI-1685 b.

However, the two-planet RV models investigated in \S\ref{sec:rv_analysis} using the combined CARMENES, IRD, and MAROON-X data sets do not support the inclusion of additional planets in the best fit RV model for this system. To confirm that our GP model is not absorbing additional signals induced by these two planet candidates, we carry out a set of tests that inject two synthetic planets into the stellar activity time series model produced by Pyaneti (teal line in the top panel of Figure \ref{fig:RV_Summary}). In one instance we inject a synthetic planet b, using the best fit orbital parameters from our fit, alongside a synthetic planet matching the candidate suggested in H21. In the second instance we insert the same synthetic planet b and a synthetic planet c matching the candidate put forth in B21. We replicate our true data sets by selecting the model RV points closest in time to the actual observations and adopting the corresponding RV errors. We then fit each data set with both a 1- and 2-planet model, using the same priors for planet b and the GP as applied in our final RV model. We include uniform priors for the second planet using the 1-sigma uncertainties from the H21 and B21 results. In both cases, comparison of the AIC values from the 1- and 2- planet models shows a clear preference for the two planet model: $\Delta$AIC = 44 in the H21 candidate case and 13 in the B21 candidate case. Thus we can be confident that if there was a second planet in the data with orbital parameters matching either of the previously published candidates, the AIC comparison would point to the two planet fit as the preferred model. As this is not the case, we rule out the Keplerian nature of the planet candidates suggested in the original B21 and H21 discovery papers. 

To investigate these signals further, we generate a Stacked Bayesian Generalized Lomb Scargle (SBGLS) periodogram using the SBGLS code\footnote{https://anneliesmortier.wordpress.com/sbgls/} presented in \citet{Mortier2017}. This approach begins by calculating a Bayesian GLS (BGLS) for the first N points of an RV time series, we adopt N=10, and then iteratively adds the next RV measurement in the series and recalculates the BGLS periodogram until the entire RV time series has been included. At this point, all of the BGLS periodograms are normalized with their respective minimum values so that one can compare how the strength of a given periodic signal grows or fades with the number of observations included.

\begin{figure}
    \centering
    \includegraphics[width=.45\textwidth]
    {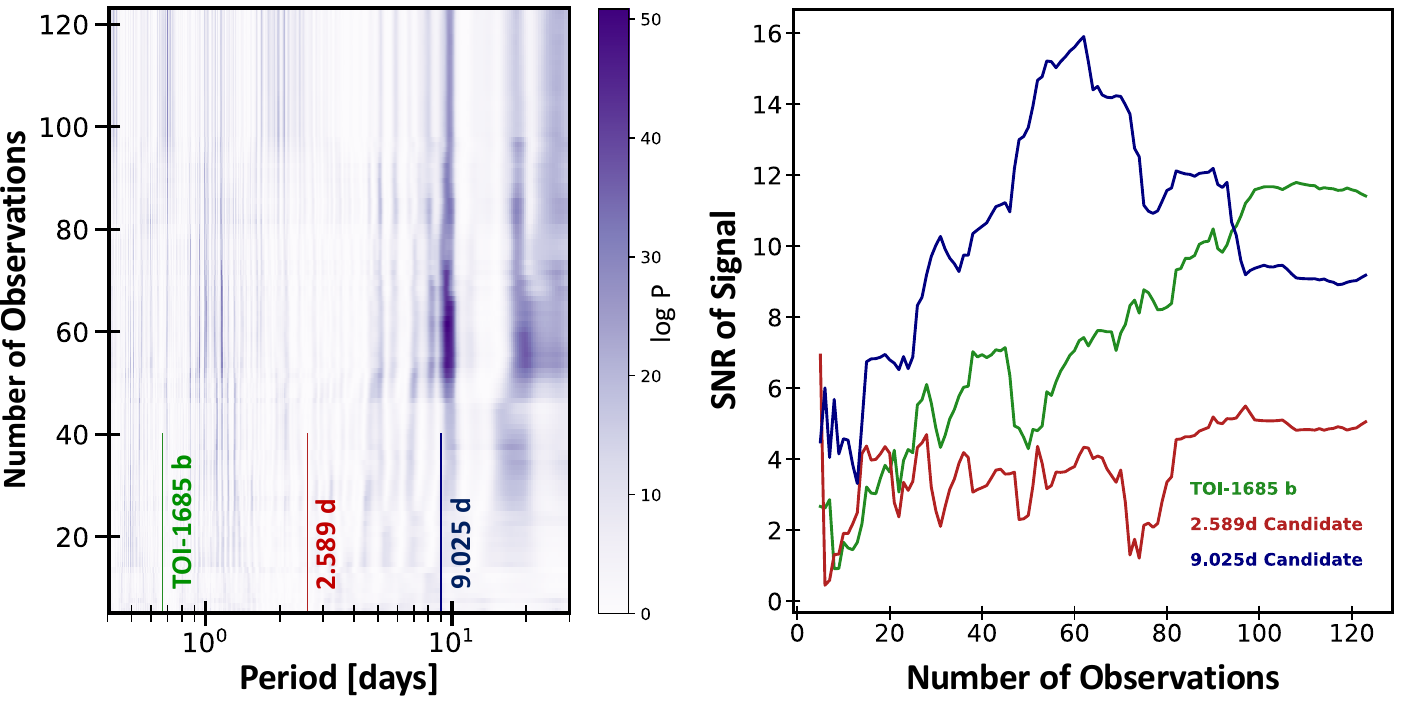}
    \caption{\textbf{Left: }Stacked Bayesian Generalized Lomb Scargle periodogram constructed using all of the 2020 RV data for TOI-1685 across the CARMENES, IRD, and MAROON-X instruments. \textbf{The vertical lines denote the periods of TOI-1685 b (green) and the potential second planets suggested in H21 (red) and in B21 (blue).} Only TOI-1685 b shows a narrow peak with a roughly monotonic increase in power as a function of the number of observations, the two hallmarks of a Keplerian signal. The H21 signal does not manifest in the combined RVs with any significant power, while the B21 signal exhibits a local maxima in the 50-70 observation range and a broad peak which suggest it is a signature of activity in the star. \textbf{Right: Detection significance of each period as a function of the number of observations. The signal at TOI-1685 b's period exhibits the roughly monotonic increase expected from a true Keplerian signal, while the signals at the periods corresponding to the H21 and B21 planet candidates show early increases in power followed by a notable downturn or leveling off as more observations are added.}}
    \label{fig:SBGLS}
\end{figure}

We only include observations taken during the 2020 observing campaigns in order to better replicate the time scale and stellar surface conditions present in the CARMENES and IRD data sets. True Keplerian signals are expected to increase their significance roughly monotonically as more data points are added and should present as narrow features over timescales that are notably longer than the planet's period thanks to their exact period repetition orbit after orbit \citep[e.g.][]{Hatzes2013, SuarezMascareno2017, Burt2021, Laliotis2023}. TOI-1685 b's  signal in the stacked periodogram meets both of these criteria, further corroborating its planetary nature (Figure \ref{fig:SBGLS}). 

The stacked periodogram does not show significant power at the 2.6 day period suggested for the planet candidate in H21, despite the higher RV precision provided by the CARMENES and MAROON-X data. H21 estimated the semi-amplitude of the signal to be $\sim$6 \ms , corresponding to a planet mass of 7-8 \Mearth. Using the publicly available \texttt{RVSearch} package \citep{Rosenthal2021}, we carry out an injection/recovery analysis of the combined RV data set and find that the existing data should be sensitive to planets down to $\sim$ 1 \Mearth\ at this 2.6 day period (Figure \ref{fig:RVSearch}). Given the lack of signal in our longer baseline and higher precision data set, we assert that the 2.6 day H21 signal is due to a non-planetary source. The true origin of the signal is not immediately obvious. It could be driven by instrumental systematics, aliases within the IRD data, or even short term variability in the star. Regardless, additional targeted investigation would be required to identify its origin.

The 9 day period of the planet candidate suggested in B21 does show significant power in the stacked periodogram, especially in the N = 50 - 70 observation range. But the power decreases as additional observations are added to the combined RV data set, making that peak a local maxima rather than a sustained detection. This characteristic of showing a probability maxima and then decreasing as additional data is added is used to rule out the Keplerian nature of a similar 19 day signal detected in the CARMENES RVs in B21 and to attribute that peak in the periodogram to stellar variability. Given the lack of sustained power increase and the relatively wide peak that the 9 day signal exhibits in the stacked periodogram (which can be suggestive of differential rotation, rather than a planet on a well defined orbit, see, e.g., \citealt{Mortier2017}) we now ascribe the 9 day signal to stellar variability as well. Significant signals at harmonics of the star's rotation (P/2, P/3, etc) are common in RV periodograms \citep[see, e.g.,][]{Boisse2011}, though this is not explicitly required and activity signals can emerge at non-rotation-related periods as well \citep{Nava2020}.

\subsection{Remaining Possibilities for a Companion Planet}

\begin{figure*}[hbt!]
    \centering
    \includegraphics[width=.9\textwidth]
    {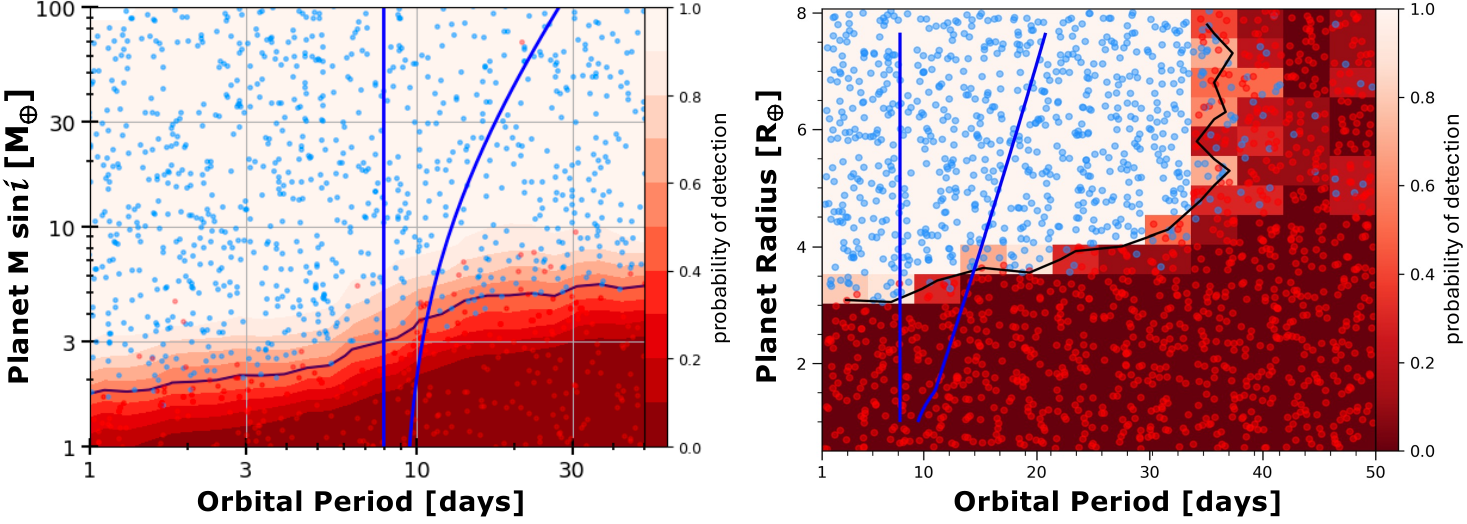}
    \caption{Injection recovery analyses of the residual RVs generated by \texttt{Pyaneti} (left) and the combined TESS S19 and S59 light curves (right). Blue and red circles denote synthetic planets that were and were not successfully recovered, respectively. The black contours denote the 50\% detection probability threshold as a function of orbital period. In both cases, small and low mass planets could evade detection, leaving the possibility that TOI-1685 b is not the only low mass planet in the system. We over plot the allowable parameter space for a perturbing planet if TOI-1685 b started on a 0.05 AU orbit as blue lines, using the mass-radius relation from \citet{Muller2023} to convert the planet masses from Figure \ref{fig:USP_perturber_parameters}. For the perturber to remain undetected, it would need to have a mass $\leq$ 3 \Mearth\ and a radius $\leq$ 3 \Rearth\ if on the shorter (8 day) end of the allowable period range. The undetected planet limits relax to $\leq$ 5 \Mearth\ and $\leq$ 3.75 \Rearth\ when considering a perturber on the longer (27 day) end of the allowed orbital period range.}
    \label{fig:RVSearch}
\end{figure*}

While these two signals do not appear to be bona fide planets that could have shepherded TOI-1685 b onto its USP orbit, there is still a range of dynamical parameter space in the system that could hold additional planets. Using the \texttt{RVSearch} python package \citep{Rosenthal2021}, we inject a population of synthetic planets into the residual RVs from Pyaneti and test what combinations of planet mass and orbital period should be detectable in the RV data sets presented in this work (Figure \ref{fig:RVSearch}). We perform a similar transit injection recovery analysis using the combined Sector 19 and Sector 59 light curves from the TESS SPOC 2-minute observations for TOI-1685. The transit signal of the known planet is masked out using its best-fit ephemeris to obtain the residual light curve. The injected planets  are drawn from a log-uniform grid in period (1-50 days) and radius  (0.5-8 Earth radii), with randomized ephemeris and impact parameters. We use the Box Least Squares (BLS) algorithm to search for the injected signal. The signal is considered ``recovered'' if the ratio of its BLS signal to the pink noise passes a threshold value of 9 and the recovered signal has a period within at least 2-sigma of the injected period using the ephemeris matching algorithms from \citealt{Coughlin:2014}.

As one summary metric we calculate the planet mass that corresponds to a 50\% detection probability on a 50 day orbit, equivalent to a semi-major axis of 0.20 AU, as \citet{Sanchis-Ojeda2014} showed that most USP super-Earth planets have companions with periods P \textless\ 50 days. With the existing RV data, \texttt{RVSearch} shows that planets with \Mpl\ $\leq$ 8\Mearth\ on 50 day orbits could go undetected. The TESS data, limited to just 56 total days of observation, is sensitive to planets down to $\sim$3\Rearth\ on periods shorter than 10 days but planets up to 8\Rearth\ would likely go undetected in a BLS search if their periods exceed 40 days. Both results thus leave open the possibility that TOI-1685 b could have a nearby, low mass, planet companion.

\subsection{Investigations into a Perturbing Companion Planet}

\begin{figure}
    \centering
    \includegraphics[width=.45\textwidth]
    {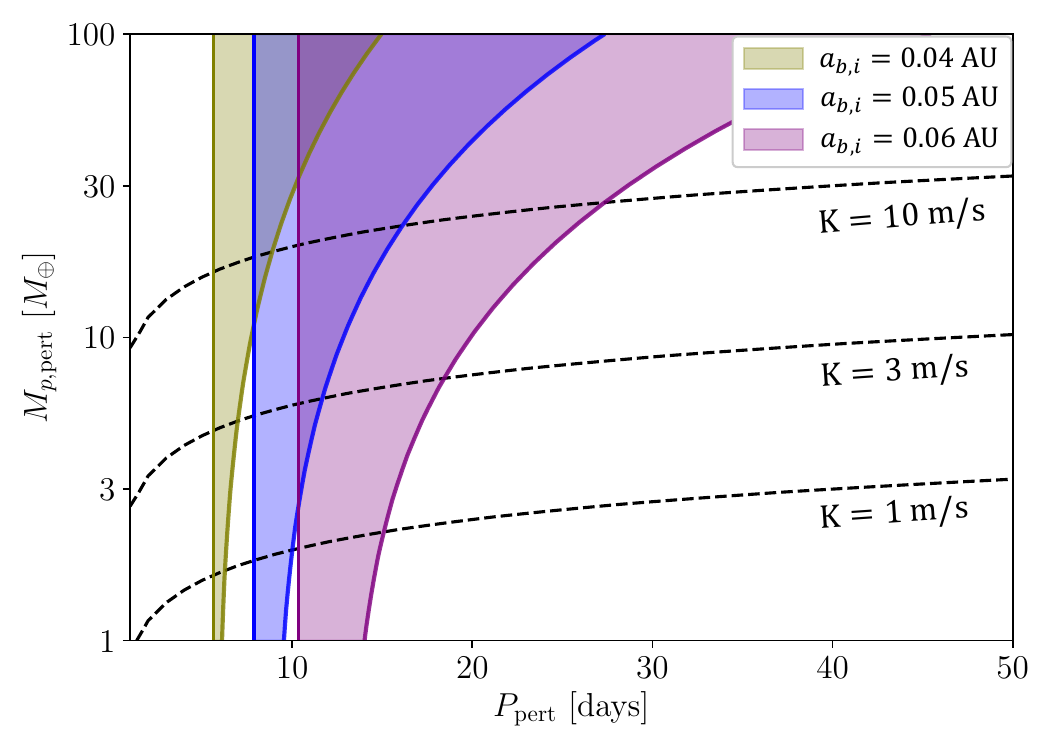}
    \caption{Period and mass of a hypothetical perturber that could have caused TOI-1685 b to decay to its current orbit via obliquity tides. The shadings indicate different possible \textit{initial} semi-major axes of TOI-1685 b before its decay. Each region is bounded on the left by a minimum initial period ratio, $P_{\mathrm{pert}}/P_{b,i} = 1.3$, and on the right by $|g|/\alpha = 1$ (with $|g|/\alpha$ increasing to the left). The dashed black lines indicate contours of constant RV semi-amplitude.}
    \label{fig:USP_perturber_parameters}
\end{figure}

Given these detection limits, it is worthwhile to determine the parameters of a hypothetical perturbing planet that could have driven TOI-1685 b's migration to its current orbit. We consider obliquity-driven tidal migration \citep{Millholland2020} for this exercise, noting that other pathways are also possible. In order for the mechanism to operate, the system needed to begin with $|g|/\alpha > 1$, where $g = \dot{\Omega}$ is the nodal precession rate of TOI-1685 b triggered by its companion and $\alpha$ is its spin-axis precession rate. We use our derived system parameters and also assume TOI-1685 b's Love number to be $k_2 = 0.2$ and moment of inertia factor to be $C = 0.3$, although the results are not strongly sensitive to these choices. In Figure \ref{fig:USP_perturber_parameters}, we plot the constraints on the period and mass of the hypothetical perturbing planet. The constraints depend on the unknown starting location of TOI-1685 b, so we consider three possible initial semi-major axes, $a_{b, i}$, at 0.04, 0.05, and 0.06 AU. We select this range (which corresponds to the semi-major axis range inhabited by the innermost planets in compact multi-planet systems) because several USP formation theories, including obliquity tides theory used here, posit that USPs started out at the inner edge of a compact multi-planet system and migrated inwards. The shaded regions indicate where the initial $|g|/\alpha > 1$ for the chosen $a_{b, i}$, and they are bounded on the right by $|g|/\alpha = 1$ and on the left by a minimum initial period ratio, $P_{\mathrm{pert}}/P_{b,i} = 1.3$. We also plot contours of the RV semi-amplitude of the hypothetical perturber.

Taking the middle option, where TOI-1685 b formed in a 0.05 AU (P $\sim$ 6 days) orbit, the perturber must have an orbital period between 7.9 and 27.4 days and a mass between 1 and 100 \Mearth, with increased mass necessary at longer periods. We outline this parameter space on the RV and transit injection/recovery grids presented in Figure \ref{fig:RVSearch} and note that for the perturber to remain undetected, it would need to have a mass $\leq$ 3 \Mearth\ and a radius $\leq$ 3 \Rearth\ if on the shorter (8 day) end of the allowable period range. The undetected planet limits relax to $\leq$ 5 \Mearth\ and $\leq$ 3.75 \Rearth\ when considering a perturber on the longer (27 day) end of the allowed orbital period range.

\subsection{Interior models}\label{sec:internal_comp}

Given the small radius, Earth-like density, and hot equilibrium temperature of the planet, we do not expect it to have retained any significant H/He atmosphere that may have accumulated during planet formation \citep{Rogers2015, Lopez2017}. The presence of a significant water layer on the planet is also unlikely, given how closely the planet falls to the Earth-like and Pure MgSiO$_{3}$ model lines on the mass-radius diagram (Figure \ref{fig:MassRadiusDiagram}) and how well its density aligns with the rocky-worlds population of the tri-modal distribution of small planets around M-dwarfs that distinguishes rocky planets from water- and gas-rich worlds \citep{Luque2022}. 

\begin{figure}[ht!]
    \includegraphics[width=.45\textwidth]{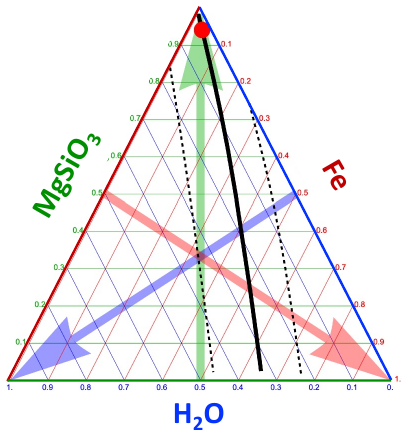}
    \caption{Ternary plot depicting different fractional combinations of water, rock and iron for solid planet compositions. The solid and dashed black lines mark the possible position of TOI-1685 b and the 68\% credible intervals. The red dot depicts the model planet at an estimated internal pressure of P$_{\mathrm{0}}$ = 600 GPa, which results in relative mass fractions of 94.2\% MgSiO$_{\rm{3}}$, 3.3\% Fe, and 2.5\% H$_{\rm{2}}$O.}
    \label{fig:InteriorComp}
\end{figure}

We use the \texttt{Manipulate Planet} tool\footnote{https://www.cfa.harvard.edu/$\sim$lzeng/manipulateplanet.html} to estimate the interior composition of TOI-1685 b using a three layer model comprised of iron, water, and magnesium silicate. To estimate the planet's central pressure, P$_{\rm{o}}$, we scale the Earth's central pressure of 600 Gpa by M$_{pl}^{2}$/R$_{pl}^{4}$, producing an estimate of P$_{\rm{o}} \approx 600$ GPa. This central pressure, combined with our best-fit measurements of the planet's mass and radius, produces an interior model dominated by Magnesium Silicate (Figure \ref{fig:InteriorComp}). Specifically, the three-component mass fractions are: 94.2\% MgSiO$_{\rm{3}}$, 3.3\% Fe, and 2.5\% H$_{\rm{2}}$O and the corresponding radius components are: R$_{\rm{MgSiO_{3}}}$ = 1.08\Rearth, R$_{\rm{Fe}}$ = 0.34\Rearth, and R$_{\rm{H_{\rm{2}}O}}$ = 0.046 \Rearth.

\subsection{Implications for atmospheric characterisation}\label{sec:atmosphere}

The updated stellar and planet parameters derived in this work impact the planet's estimated potential for atmospheric characterization. The standard metrics for assessing this potential are the Transmission and Emission Spectroscopy Metrics (TSM and ESM), defined in Equations 1 and 4 of \citet{Kempton:2018}. The TSM is proportional to the transmission spectroscopy signal of a planet based on the expected strength of its spectral features and the J-band magnitude of the host star, while the ESM is proportional to the secondary eclipse signal at mid-IR wavelengths based on the planet's size and temperature, and the K-band magnitude of the host star. At present, TOI-1685 b will be targeted by three separate \JWST\ programs: two transmission spectroscopy programs using the NIRSPEC/BOTS \citep{Jakobsen2022} and NIRISS/SOSS \citep{Doyon2012} configurations, and one thermal emission program using NIRSPEC/BOTS. 

Inserting our new planet and stellar parameters results in an updated TSM value of 12.4 $\pm$ 2.0 which just barely surpasses the TSM = 12 cut off suggested for identifying the best \JWST\ targets smaller than 1.5 \Rearth\ \citep{Kempton:2018}. However, for the accepted NIRSPEC observations \citep{Fisher_proposal,Luque_proposal} the fact that TOI-1685 is an M dwarf and therefore brighter in K band than in J band (J - K$_{\rm{s}}$ = 0.858 magnitudes) bodes well as the G395 band of NIRSpec starts at 2.8$\mu$ and the K band's central wavelength of $\sim$2.2$\mu$ better approximates the star's G395 brightness than the J band's 1.25$\mu$ center. 

Updating the ESM value of TOI-1685 b using our new planet and stellar parameters gives ESM = 11.9 $\pm$ 1.3. This exceeds the cutoff value of 7.5 suggested for identifying the top emission spectroscopy small planets for \JWST\ in \citet{Kempton:2018}, and is a promising starting point for the thermal emission observations.

In order to accurately constrain the atmospheric properties of any exoplanet, one requires a certain precision on the mass and radius. Several studies have attempted to quantify the precision requirement on the planet mass, in particular \cite{Batalha2019} concluded that a $\pm50\%$ mass precision is sufficient for initial characterization with \JWST, whilst $\pm20\%$ is required for more detailed analyses. At the $\pm20\%$ level, the width of the posterior distributions, from an atmospheric retrieval analysis for example, are dominated by the observational uncertainties. In contrast, \cite{diMiao2023} determined that a mass precision of just $\pm50\%$ was sufficient to obtain reliable retrieval results, even in the case of a high mean molecular weight atmosphere. However, this study was done in the context of Ariel spectra, and relies on the assumption that one can accurately retrieve the planet mass from a spectroscopic analysis, which may be challenging due to degeneracies between mass and composition \citep{Batalha2017}. Our new constraints on the mass of TOI-1685 b provide a precision of $\sim\pm10\%$, which will enable a detailed characterization of any potential atmosphere on this planet. 

Furthermore, the scale height of the atmosphere, which has a significant effect on the size of all the spectral features, is given by 
\begin{equation}
    H = \frac{k_B T}{\mu g},
\end{equation}
where $T$ is the atmospheric temperature, $\mu$ is the mean molecular weight, $g$ is the surface gravity, and $k_B$ is the Boltzmann constant. This leads to a direct degeneracy between the surface gravity, determined through the planetary mass and radius, and the mean molecular weight of the atmosphere. A key science goal of \JWST\ is to establish the presence and nature of the atmospheres of smaller planets, which are unlikely to be hydrogen-dominated. In the case of TOI-1685 b, future \JWST\ observations hope to observe its possible atmosphere, and measuring the mean molecular weight will provide some constraints on its chemical composition. Our updated mass and radius give a value for the surface gravity of $1378^{+262}_{-226}$ cm s$^{-2}$, a precision of $<20\%$. This should enable one to determine $\mu$ from the transmission spectra, and distinguish between different atmospheric scenarios. 

\section{Conclusion}\label{sec:conclusion}

Here we present an updated analysis of the small, ultra short period planet orbiting the nearby M-dwarf star TOI-1685. This planet was first identified by NASA's \TESS\ mission, and the host star was then followed up photometrically from the ground by the MuSCAT telescope network, the OMM telescope's PESTO camera, and LCO's Sinistro camera which provided additional transit photometry used to refine the planet's radius. The star has also been observed by the CARMENES, IRD, and MAROON-X spectrographs which provide precise RV measurements that are used to determine the planet's mass.

We highlight the importance of the stellar characterization phase of any exoplanet confirmation effort, as study of the star's photometric metallicity reveals that it lies 0.57 magnitudes above the main sequence and must therefore be metal-rich ([Fe/H] = +0.3 $\pm$ 0.1). This increased stellar metallicity affects the star's mass and radii estimates as derived using the \citet{Mann2015} and \citet{Mann2019} absolute magnitude-metallicity- and mass / radius calibrations. This in turn impacts the planet's derived mass and radius values, as these are indirect measurements and report the planet's mass and radius relative to its host star. Our updated stellar and planetary mass and radius values are as follows: \Rstar = 0.4555$\pm$ 0.0128 \Rsun, \Mstar = 0.454 $\pm$ 0.018 \Msun, R$_{\rm{p}}$ = 1.468$^{+0.050}_{-0.051}$ \Rearth, and M$_{\rm{p}}$ = 3.03$^{+0.33}_{-0.32}$ \Mearth. Together, the planetary mass and radius produce a best fit bulk density value of $\rho_{\rm{p}}$ = 5.3 $\pm$ 0.8 g cm$^{-3}$, just 4\% less than the Earth's bulk density of 5.51 g cm$^{-3}$. Combining our updated Luminosity (\Lstar\ = 0.03052 $\pm$ 0.00061 \Lsun) and effective temperature (\Teffstar\ = 3575 $\pm$ 53 K) estimates for the star with the planet's orbital period and assumed Bond Albedo of 0.1 produces an estimated equilibrium temperature of \teq\ = 1062 $\pm$ 27 K for TOI-1685 b.

Our updated transit and RV analyses suggest that TOI-1685 falls between the planet solutions presented in the B21 and H21 discovery papers (Figure \ref{fig:MassRadiusDiagram}). The planet is smaller and less massive than suggested in B21 (with the radius decreasing by 15.8\% and the mass decreasing by 24.8\%) and roughly the same size but less massive than the best fit results from H21 (with the radius increasing by 0.6\% and the mass decreasing by 13.2\%). Follow up efforts to confirm and obtain masses for transiting planets are often required to pass a mass precision threshold before being considered ready for publication (3-$\sigma$ and 5-$\sigma$ being common break points). There is thus a possibility that planets' whose RV semi-amplitudes are measured to be slightly too high are more likely to get published than planets whose RV semi-amplitudes are measured to be slightly too low. This can influence which planets appear in the literature and are therefore included on the low-mass end of exoplanet mass-radius diagrams, and may then bias our mass-radius relations \citep[see, e.g.,][]{Burt2018, Montet2018}. While time intensive, it is nonetheless important to continue RV monitoring of small planets past their initial publication to minimize such biases.

We do not find sufficient evidence to confirm the planetary nature of the additional planet candidates put forth in the H21 and B21 publications. The 9 day signal in B21 appears prominently in the combined RV data set, but is likely a P$_{\rm{rot}}$/2 manifestation of the star's rotational modulation based upon a Stacked Bayesian Generalized Lomb Scargle analysis. The 2.6 day signal suggested in H21 does not present at a significant level in the combined RV data set, despite the increased observational baseline and RV precision, and so we assume it is a systematic tied to the original IRD data. A set of injection / recovery analyses applied to the RV and TESS data, however, reveal that small (\Rpl\ \textless 3.75 \Rearth) and low mass (\Mpl\ \textless 5 \Mearth) planets could still remain hidden in the existing data on orbits less than 30 days.

When investigating the parameters of a hypothetical second, perturbing planet that could have driven TOI-1685 b's migration to its current orbit via obliquity tides, we find that if TOI-1685 b formed at a separation of 0.05 AU (P $\sim$ 6 days) then the perturber must have an orbital period between 7.9 and 27.4 days and a mass from 1 - 100 \Mearth\ with increased mass necessary at longer periods. For such a planet to remain undetected, it would need to have a mass $\leq$ 3 \Mearth\ and a radius $\leq$ 3 \Rearth\ if on the shortest end of that period range. The planet could, however, be as massive as 5 \Mearth\ and as large as 3.75 \Rearth\ and still avoid detection in the existing RV and photometry data if on the longest end of the allowed period range.

Obtaining planet masses and radii that are both precise \textit{and} accurate is especially crucial in the era of \TESS\ \& \JWST. As noted in the beginning of this paper, TOI-1685 b is the subject of three accepted \JWST\ proposals scheduled for observation in Cycle 2, two of which focus on the planet's status as a bona fide water world while the third aims to measure its oxidation as a hot, rocky, super-Earth. Our new mass and radius measurements deem the water world scenario unlikely, but a high mean-molecular weight atmosphere is still plausible, though more challenging to observe. Even in the absence of any atmosphere, however, the phase curve measurements could still provide information about the planet's interior composition. In any case, our improved mass and radius constraints will be essential for correctly interpreting these upcoming observations.

The \TESS\ mission continues to produce thousands of tantalizing small exoplanet candidates, far more than can be thoroughly confirmed and characterized by the limited number of over-subscribed high precision RV spectrographs currently in operation. Ensuring that the TOIs most likely to be included in the \textit{even more} over-subscribed \JWST\ schedule have undergone sufficient follow up and characterization (both of the planets and of their host stars) to know they are truly appropriate targets for the stated \JWST\ science goals should be a primary focus of the exoplanet community moving forward.

\section*{Acknowledgements}

Part of this research was carried out at the Jet Propulsion Laboratory, California Institute of Technology, under a contract with the National Aeronautics and Space Administration (NASA). E.G. was supported by NASA NSF Astronomy \& Astrophysics Research Program Grant 1817215, NASA Exoplanets Research Program Grant 80NSSC20K0251, and NASA \TESS\ Guest Observer Cycle 4 Award 80NSSC22K0295. The University of Chicago group acknowledges funding for the MAROON-X project from the David and Lucile Packard Foundation, the Heising-Simons Foundation, the Gordon and Betty Moore Foundation, the Gemini Observatory, the NSF (award number 2108465), and NASA (grant number 80NSSC22K0117). 

We follow the \dataset[guidelines]{https://doi.org/10.5281/zenodo.10161527} of NASA’s Transform to OPen Science (TOPS) mission for our open science practices.

This paper includes data collected by the \TESS\ mission. Funding for the \TESS\ mission is provided by the NASA Explorer Program. Resources supporting this work were provided by the NASA High-End Computing (HEC) Program through the NASA Advanced Supercomputing (NAS) Division at Ames Research Center for the production of the SPOC data products.

The TESS data presented in this article were obtained from the Mikulski Archive for Space Telescopes (MAST) at the Space Telescope Science Institute and can be accessed via this \dataset[DOI]{https://doi.org/10.17909/h9bv-4e03}. Support for MAST for non-\HST\ data is provided by the NASA Office of Space Science via grant NNX13AC07G and by other grants and contracts.

This work was enabled by observations made from the Gemini North telescope, located within the Maunakea Science Reserve and adjacent to the summit of Maunakea. We are grateful for the privilege of observing the universe from a place that is unique in both its astronomical quality and its cultural significance. The Gemini observations presented in this work are associated with programs GN-2020B-Q-234 and GN-2021B-Q-230. 

The software used to execute most analyses in this paper and to generate the corresponding figures is hosted at https://github.com/JenniferBurt /TOI-1685b and is preserved, alongside the input data sets, on Zenodo at this \dataset[DOI]{10.5281/zenodo.11105468}

This research has made use of the NASA Exoplanet Archive, which is operated by the California Institute of Technology, under contract with the National Aeronautics and Space Administration under the Exoplanet Exploration Program. This research has also made use of the Exoplanet Follow-up Observation Program website, which is operated by the California Institute of Technology, under contract with the National Aeronautics and Space Administration under the Exoplanet Exploration Program. 

This research has made use of NASA’s Astrophysics Data System Bibliographic Services.

This research made use of Astropy, a community-developed core Python package for Astronomy \citep{Astropy2013,Astropy2018,Astropy2022}. 

Based on observations obtained with PESTO at the Mont-Mégantic Observatory, funded by the Université de Montréal, Université Laval, the Natural Sciences and Engineering Research Council of Canada (NSERC), the Fond québécois de la recherche sur la Nature et les technologies (FQRNT) and the Canada Economic Development program.

We thank Li Zeng for guidance on the ManipulatePlanet tool. We thank Akihiko Fukui, John Livingston, and the MuSCAT team for sharing the MuSCAT light curves originally presented in \citet{Hirano2020} for use in this work. We also thank Hannu Parviainen and the MuSCAT2 team for sharing a custom reduction of all of the available MuSCAT2 observations of TOI-1685 and advise on their use.

We also thank the anonymous referee for their time and care in reviewing this paper. Their suggestions improved the clarity, depth, and utility of this work.

\facilities{TESS, Okayama-1.88m/MuSCAT,\\ TCS/MuSCAT2, FTN/MuSCAT3, LCO/SINISTRO, OMM/PESTO, Gemini North/MAROON-X,\\ Subaru/IRD, Centro Astronómico Hispano-Alemán/CARMENES} 

\software{\texttt{ARIADNE} \citep{Vines2022}, 
\texttt{Astropy} \citep{Astropy2022},
\texttt{BANYAN $\Sigma$} \citep{Gagne2018}, 
\texttt{batman} \citep{Kreidberg2015}, 
\texttt{celerite} \citep{ForemanMackey2017}, 
\texttt{dynesty} \citep{Speagle2020}, 
\texttt{juliet} \citep{Espinoza2019},
\texttt{Manipulate Planet} \citep{Zeng2016},
\texttt{Pyaneti} \citep{Barragan2019, Barragan2022}, 
\texttt{RVSearch} \citep{Rosenthal2021}, 
\texttt{SBGLS} \citep{Mortier2017}, 
\texttt{SERVAL} \citep{Zechmeister2018}, 
\texttt{VOSA} \citep{Bayo2008}.}

\bibliographystyle{apj}
\bibliography{refs}



\begin{appendix} 

\clearpage


\section{Transit model limb darkening, GP, baseline and noise parameters}
\begin{table}[h]
\centering
\caption{Best-fitting values for the limb darkening, GP, baseline and noise parameters in the transit model described in \S\ref{sec:transit_phot}. The auxiliary parameters used to detrend each light curve are summarised in Table \ref{tab:ground}, and the associated coefficient in the table represents the best fitting value when each vector is normalised to have a maximum of 1 and a minimum of -1.}
\label{tab:noise}
\begin{tabular}{lc||lc||lc}
\hline \hline
Parameter [Unit] & Value & Parameter [Unit] & Value & Parameter [Unit] & Value \\ 
\hline 
$q_\mathrm{1,TESS}$ & $0.48^{+0.2}_{-0.19}$ &  $\delta y_\mathrm{M12z}$ & $-0.0007\pm0.00022$ &  $am_\mathrm{M22z}$ & $0.0\pm0.00019$\\ 
$q_\mathrm{2,TESS}$ & $0.161^{+0.11}_{-0.097}$ &  $FWHM_\mathrm{M12z}$ & $-0.00006^{+0.00022}_{-0.00023}$ &  $\delta x_\mathrm{M22z}$ & $-0.00001^{+0.0003}_{-0.00029}$\\ 
$q_{1,g^\prime}$ & $0.25^{+0.16}_{-0.14}$ &  $peak_\mathrm{M12z}$ & $-0.0001\pm0.00019$ &  $\delta y_\mathrm{M22z}$ & $-0.0\pm0.00027$\\ 
$q_{2,g^\prime}$ & $0.74^{+0.16}_{-0.2}$ &  $c_\mathrm{M13g}$ & $-0.00305^{+0.00031}_{-0.0003}$ &  $FWHM_\mathrm{M22z}$ & $0.00008\pm0.00029$\\ 
$q_{1,r^\prime}$ & $0.83^{+0.11}_{-0.12}$ &  $\sigma_\mathrm{w,M13g}$ [ppm] & $24^{+110}_{-20}$ &  $c_\mathrm{M31g}$ & $-0.00012^{+0.00024}_{-0.00025}$\\ 
$q_{2,r^\prime}$ & $0.82^{+0.1}_{-0.12}$ &  $am_\mathrm{M13g}$ & $0.00446^{+0.00047}_{-0.00048}$ &  $\sigma_\mathrm{w,M31g}$ [ppm] & $46^{+120}_{-35}$\\ 
$q_{1,i^\prime}$ & $0.57^{+0.15}_{-0.14}$ &  $\delta x_\mathrm{M13g}$ & $-0.00008^{+0.00043}_{-0.00042}$ &  $am_\mathrm{M31g}$ & $0.00101^{+0.0002}_{-0.00019}$\\ 
$q_{2,i^\prime}$ & $0.7^{+0.17}_{-0.2}$ &  $\delta y_\mathrm{M13g}$ & $0.00112^{+0.00037}_{-0.00036}$ &  $\delta x_\mathrm{M31g}$ & $0.00028^{+0.00032}_{-0.00033}$\\ 
$q_{1,z_s}$ & $0.27^{+0.15}_{-0.13}$ &  $FWHM_\mathrm{M13g}$ & $-0.00089^{+0.00059}_{-0.00058}$ &  $\delta y_\mathrm{M31g}$ & $0.00011\pm0.00023$\\ 
$q_{2,z_s}$ & $0.75^{+0.16}_{-0.2}$ &  $peak_\mathrm{M13g}$ & $-0.00178\pm0.0008$ &  $FWHM_\mathrm{M31g}$ & $-0.00021\pm0.00046$\\ 
$GP_\mathrm{S0,1}/10^{10}$ & $9.3^{+23.0}_{-5.8}$ &$c_\mathrm{M13z}$ & $0.0002\pm0.00022$ &  $peak_\mathrm{M31g}$ & $-0.00122^{+0.00039}_{-0.0004}$\\ 
$GP_\mathrm{P,1}$ [d] & $4.9^{+6.7}_{-3.7}$ &  $\sigma_\mathrm{w,M13z}$ [ppm] & $250^{+170}_{-200}$ &  $c_\mathrm{M31r}$ & $-0.00073^{+0.00025}_{-0.00024}$\\ 
$GP_\mathrm{S0,2}/10^{10}$ & $34^{+22}_{-12}$ &  $am_\mathrm{M13z}$ & $-0.00078^{+0.00019}_{-0.0002}$ &  $\sigma_\mathrm{w,M31r}$ [ppm] & $160^{+190}_{-100}$\\ 
$GP_\mathrm{P,2}$ [d] & $3.3^{+1.2}_{-1.0}$ &  $\delta x_\mathrm{M13z}$ & $-0.00012\pm0.00014$ &  $am_\mathrm{M31r}$ & $0.00007\pm0.00011$\\ 
$c_\mathrm{TESS19}$ & $-0.000071^{+0.000032}_{-0.000031}$ &  $\delta y_\mathrm{M13z}$ & $-0.00019\pm0.00016$ &  $\delta x_\mathrm{M31r}$ & $0.00009^{+0.00022}_{-0.00023}$\\ 
$\sigma_\mathrm{w,TESS19}$ [ppm] & $16^{+26}_{-11}$ &  $FWHM_\mathrm{M13z}$ & $0.00079\pm0.00025$ &  $\delta y_\mathrm{M31r}$ & $0.00019\pm0.00018$\\ 
$c_\mathrm{TESS59}$ & $-0.000071^{+0.000047}_{-0.000049}$ &  $peak_\mathrm{M13z}$ & $0.00007^{+0.00023}_{-0.00022}$ &  $FWHM_\mathrm{M31r}$ & $0.00045^{+0.0003}_{-0.00029}$\\ 
$\sigma_\mathrm{w,TESS59}$ [ppm] & $34^{+75}_{-26}$ &  $c_\mathrm{M21r}$ & $-0.0005^{+0.00054}_{-0.00053}$ &  $peak_\mathrm{M31r}$ & $0.00065\pm0.00026$\\ 
$c_\mathrm{M11z}$ & $0.00019\pm0.00024$ &  $\sigma_\mathrm{w,M21r}$ [ppm] & $9944^{+39}_{-65}$ &  $c_\mathrm{M31i}$ & $-0.00061^{+0.00023}_{-0.00022}$\\ 
$\sigma_\mathrm{w,M11z}$ [ppm] & $11.2^{+110.0}_{-9.0}$ &  $am_\mathrm{M21r}$ & $0.00105^{+0.00077}_{-0.00076}$ &  $\sigma_\mathrm{w,M31i}$ [ppm] & $5.9^{+20.0}_{-4.1}$\\ 
$am_\mathrm{M11z}$ & $-0.000594^{+0.000096}_{-0.000098}$ &  $\delta x_\mathrm{M21r}$ & $0.002\pm0.0013$ &  $am_\mathrm{M31i}$ & $-0.000284\pm0.000084$\\ 
$\delta x_\mathrm{M11z}$ & $0.00052\pm0.00017$ &  $\delta y_\mathrm{M21r}$ & $0.0015^{+0.00068}_{-0.0007}$ &  $\delta x_\mathrm{M31i}$ & $0.00013\pm0.00014$\\ 
$\delta y_\mathrm{M11z}$ & $-0.00001\pm0.0002$ &  $FWHM_\mathrm{M21r}$ & $-0.0001\pm0.0013$ &  $\delta y_\mathrm{M31i}$ & $-0.00017\pm0.00015$\\ 
$FWHM_\mathrm{M11z}$ & $-0.00078\pm0.00033$ &  $c_\mathrm{M21i}$ & $-0.00035^{+0.00031}_{-0.0003}$ &  $FWHM_\mathrm{M31i}$ & $0.00071^{+0.00025}_{-0.00026}$\\ 
$peak_\mathrm{M11z}$ & $-0.00135\pm0.00029$ &  $\sigma_\mathrm{w,M21i}$ [ppm] & $5690^{+170}_{-160}$ &  $peak_\mathrm{M31i}$ & $0.00101\pm0.00023$\\ 
$c_\mathrm{M12g}$ & $-0.00477^{+0.00029}_{-0.00028}$ &  $am_\mathrm{M21i}$ & $0.00073^{+0.00048}_{-0.00051}$ &  $c_\mathrm{M31z}$ & $0.0^{+0.00022}_{-0.00023}$\\ 
$\sigma_\mathrm{w,M12g}$ [ppm] & $17^{+66}_{-13}$ &  $\delta x_\mathrm{M21i}$ & $0.00071^{+0.00072}_{-0.00074}$ &  $\sigma_\mathrm{w,M31z}$ [ppm] & $18^{+47}_{-14}$\\ 
$am_\mathrm{M12g}$ & $0.00726^{+0.00035}_{-0.00037}$ &  $\delta y_\mathrm{M21i}$ & $0.00048\pm0.00039$ &  $am_\mathrm{M31z}$ & $-0.000433^{+0.000096}_{-0.000094}$\\ 
$\delta x_\mathrm{M12g}$ & $0.00055\pm0.00045$ &  $FWHM_\mathrm{M21i}$ & $0.00005^{+0.0008}_{-0.00079}$ &  $\delta x_\mathrm{M31z}$ & $0.00064^{+0.0002}_{-0.00019}$\\ 
$\delta y_\mathrm{M12g}$ & $0.00018^{+0.00053}_{-0.00051}$ &  $c_\mathrm{M21z}$ & $-0.00032^{+0.00029}_{-0.00028}$ &  $\delta y_\mathrm{M31z}$ & $-0.00018\pm0.00016$\\ 
$FWHM_\mathrm{M12g}$ & $-0.0003\pm0.00039$ &  $\sigma_\mathrm{w,M21z}$ [ppm] & $4490\pm140$ &  $FWHM_\mathrm{M31z}$ & $0.00021\pm0.00025$\\ 
$peak_\mathrm{M12g}$ & $-0.00163^{+0.00041}_{-0.00039}$ &  $am_\mathrm{M21z}$ & $0.00073\pm0.00039$ &  $peak_\mathrm{M31z}$ & $-0.00035\pm0.00023$\\ 
$c_\mathrm{M12r}$ & $-0.00065^{+0.00022}_{-0.00021}$ &  $\delta x_\mathrm{M21z}$ & $0.00062^{+0.00058}_{-0.00059}$ &  $c_\mathrm{OMM}$ & $0.00183^{+0.00032}_{-0.00033}$\\ 
$\sigma_\mathrm{w,M12r}$ [ppm] & $130^{+150}_{-81}$ &  $\delta y_\mathrm{M21z}$ & $0.00019^{+0.00031}_{-0.00029}$ &  $\sigma_\mathrm{w,OMM}$ [ppm] & $1331^{+97}_{-95}$\\ 
$am_\mathrm{M12r}$ & $0.00095\pm0.00012$ &  $FWHM_\mathrm{M21z}$ & $-0.00044^{+0.00069}_{-0.00065}$ &  $am_\mathrm{OMM}$ & $0.00019\pm0.00013$\\ 
$\delta x_\mathrm{M12r}$ & $-0.00029\pm0.00024$ &  $c_\mathrm{M22i}$ & $0.00001^{+0.00022}_{-0.00023}$ &  $bg_\mathrm{OMM}$ & $0.0022^{+0.00032}_{-0.00033}$\\ 
$\delta y_\mathrm{M12r}$ & $-0.00037\pm0.00019$ &  $\sigma_\mathrm{w,M22i}$ [ppm] & $2165^{+86}_{-79}$ &  $c_\mathrm{LCO1}$ & $0.00036\pm0.00021$\\ 
$FWHM_\mathrm{M12r}$ & $-0.00071\pm0.00023$ &  $am_\mathrm{M22i}$ & $0.00035\pm0.00027$ &  $\sigma_\mathrm{w,LCO1}$ [ppm] & $903^{+95}_{-90}$\\ 
$peak_\mathrm{M12r}$ & $-0.00055\pm0.00022$ &  $\delta x_\mathrm{M22i}$ & $-0.00011\pm0.00032$ &  $am_\mathrm{LCO1}$ & $-0.00083\pm0.00015$\\ 
$c_\mathrm{M12z}$ & $0.00028\pm0.00021$ &  $\delta y_\mathrm{M22i}$ & $-0.00005^{+0.00039}_{-0.00038}$ &  $c_\mathrm{LCO2}$ & $-0.0\pm0.00019$\\ 
$\sigma_\mathrm{w,M12z}$ [ppm] & $21^{+77}_{-17}$ &  $FWHM_\mathrm{M22i}$ & $-0.00007^{+0.00032}_{-0.00033}$ &  $\sigma_\mathrm{w,LCO2}$ [ppm] & $740^{+65}_{-60}$\\ 
$am_\mathrm{M12z}$ & $-0.00063^{+0.00012}_{-0.00011}$ &  $c_\mathrm{M22z}$ & $0.00005\pm0.00021$ &  $bg_\mathrm{LCO2}$ & $0.00028\pm0.0002$\\ 
$\delta x_\mathrm{M12z}$ & $-0.00007\pm0.00018$ &  $\sigma_\mathrm{w,M22z}$ [ppm] & $2033^{+70}_{-68}$ &  \\
\hline\hline
\end{tabular} 
\end{table}

\section{Radial Velocity Data Set}

\begin{table*}
\begin{center}
  \caption{Combined RV data set used in the Pyaneti fit described in \S\ref{sec:rv_analysis}. Instruments include CARMENES (CAR) and IRD, both of which observed the star over a single semester in 2020, and MAROON-X which observed the star across two semesters in 2020 and 2021. The MAROON-X instrument labels contain information on which detector the RV is derived from (Blue or Red, represented as B and R) and which year the observation was taken during (2020 or 2021) as each combination is treated as a separate instrument in the RV fit.}  
  \label{tab:RV_Data}
  \footnotesize
  \begin{tabular}{llll||llll||llll}
  \hline
  \hline
  \noalign{\smallskip}
  Date &	RV & $\sigma_{RV}$ & Inst & Date &	RV & $\sigma_{RV}$ & Inst &	Date & RV &	$\sigma_{RV}$ & Inst \\

  [BJD - 2450000] & [\ms] & [\ms] & & [BJD - 2450000] & [\ms] & [\ms] & & [BJD - 2450000] & [\ms] & [\ms] & \\
  \noalign{\smallskip}
  \hline
  \noalign{\smallskip}
9069.6744	&	-1.87	&	2.43	&	CAR	&	9153.02323	&	6.14	&	3.89	&	IRD	&	9184.07505	&	2.52	&	1.05	&	MXR\_20	\\
9073.6703	&	8.05	&	4.91	&	CAR	&	9153.03399	&	8.67	&	3.62	&	IRD	&	9185.03484	&	-2.15	&	1.12	&	MXB\_20	\\
9095.6713	&	-2.74	&	2.14	&	CAR	&	9153.382	&	9.91	&	2.4	&	CAR	&	9185.03484	&	-1.64	&	0.65	&	MXR\_20	\\
9097.6749	&	14.2	&	2.81	&	CAR	&	9153.4717	&	9.87	&	2.05	&	CAR	&	9188.86894	&	4.06	&	3.73	&	IRD	\\
9098.6754	&	5.97	&	1.78	&	CAR	&	9154.4992	&	4.88	&	1.76	&	CAR	&	9188.87963	&	7.77	&	5.12	&	IRD	\\
9099.6693	&	-1.93	&	2.13	&	CAR	&	9154.6201	&	7.08	&	1.73	&	CAR	&	9188.95689	&	-2.3	&	3.06	&	IRD	\\
9101.6859	&	-0.34	&	2.63	&	CAR	&	9156.4518	&	1.17	&	1.6	&	CAR	&	9188.9711	&	-0.28	&	3.46	&	IRD	\\
9102.6845	&	-15.14	&	3.71	&	CAR	&	9156.5853	&	3.51	&	1.61	&	CAR	&	9188.98531	&	2.51	&	3.05	&	IRD	\\
9103.6777	&	1.44	&	1.84	&	CAR	&	9156.94231	&	-11.49	&	3.8	&	IRD	&	9189.79771	&	0.93	&	3.18	&	IRD	\\
9114.7106	&	10.09	&	3.26	&	CAR	&	9156.95303	&	-8.94	&	3.82	&	IRD	&	9189.8084	&	-4.72	&	3.18	&	IRD	\\
9118.6966	&	0.46	&	2.46	&	CAR	&	9157.02692	&	-10.68	&	3.84	&	IRD	&	9189.86777	&	-6.05	&	3.13	&	IRD	\\
9119.9896	&	5.2	&	3.27	&	IRD	&	9157.0376	&	-7.3	&	3.81	&	IRD	&	9189.87845	&	-4.89	&	3.08	&	IRD	\\
9120.00035	&	7.13	&	3.66	&	IRD	&	9161.3599	&	8.37	&	3.26	&	CAR	&	9242.89031	&	-2.23	&	4.1	&	IRD	\\
9120.6746	&	-7.34	&	1.95	&	CAR	&	9161.4505	&	9.47	&	3.97	&	CAR	&	9242.90099	&	-1.21	&	4.08	&	IRD	\\
9121.6352	&	-10.46	&	1.9	&	CAR	&	9161.5732	&	0.63	&	1.63	&	CAR	&	9242.91167	&	-0.68	&	4.08	&	IRD	\\
9122.6841	&	-0.98	&	1.82	&	CAR	&	9161.6724	&	-4.24	&	2.18	&	CAR	&	9242.92235	&	0.44	&	4.23	&	IRD	\\
9123.01542	&	-2.98	&	3.11	&	IRD	&	9163.3774	&	5.92	&	2.99	&	CAR	&	9242.93303	&	-4.34	&	3.77	&	IRD	\\
9123.02619	&	-8.46	&	3.17	&	IRD	&	9163.4992	&	-1.73	&	3.03	&	CAR	&	9242.94373	&	3.64	&	4.04	&	IRD	\\
9123.03702	&	-8.64	&	3.15	&	IRD	&	9173.89166	&	-5.91	&	1.95	&	MXB\_20	&	9244.83737	&	-5.21	&	3.99	&	IRD	\\
9127.6877	&	-1.46	&	2.61	&	CAR	&	9173.89166	&	-2.82	&	1.19	&	MXR\_20	&	9244.84805	&	-2.1	&	3.78	&	IRD	\\
9128.6223	&	2.87	&	1.83	&	CAR	&	9173.98062	&	-3.88	&	1.5	&	MXB\_20	&	9244.85875	&	-5.52	&	3.76	&	IRD	\\
9131.6723	&	1.86	&	1.81	&	CAR	&	9173.98062	&	1.55	&	0.96	&	MXR\_20	&	9244.92807	&	1.59	&	3.76	&	IRD	\\
9132.6702	&	10.82	&	2.8	&	CAR	&	9180.8736	&	2.3	&	1.28	&	MXB\_20	&	9244.93875	&	4.65	&	3.8	&	IRD	\\
9138.6485	&	5.9	&	3.33	&	CAR	&	9180.8736	&	1.96	&	0.82	&	MXR\_20	&	9247.74093	&	-5.18	&	4.88	&	IRD	\\
9139.4464	&	-5.21	&	2.91	&	CAR	&	9180.9591	&	-3	&	1.26	&	MXB\_20	&	9247.7495	&	-3.09	&	4.6	&	IRD	\\
9139.5469	&	-7.91	&	1.66	&	CAR	&	9180.9591	&	0.66	&	0.82	&	MXR\_20	&	9247.75807	&	-9.92	&	4.68	&	IRD	\\
9139.6241	&	-4.77	&	2.93	&	CAR	&	9181.03145	&	-1.31	&	1.49	&	MXB\_20	&	9247.76665	&	-11.33	&	5.29	&	IRD	\\
9139.7292	&	-5.76	&	1.91	&	CAR	&	9181.03145	&	-3.8	&	0.9	&	MXR\_20	&	9515.03854	&	-8.55	&	1.18	&	MXB\_21	\\
9140.5196	&	-4.64	&	1.54	&	CAR	&	9181.84025	&	1.5	&	1.8	&	MXB\_20	&	9515.03854	&	-6	&	0.76	&	MXR\_21	\\
9140.5965	&	-3.62	&	1.65	&	CAR	&	9181.84025	&	-1.03	&	1.09	&	MXR\_20	&	9520.12791	&	-0.04	&	1.37	&	MXB\_21	\\
9140.6963	&	-3.5	&	1.72	&	CAR	&	9181.93144	&	4.34	&	1.34	&	MXB\_20	&	9520.12791	&	-1.66	&	0.79	&	MXR\_21	\\
9141.5171	&	-11.99	&	2.8	&	CAR	&	9181.93144	&	1.48	&	0.85	&	MXR\_20	&	9525.8648	&	11.91	&	1.77	&	MXB\_21	\\
9141.5792	&	-9.55	&	2.78	&	CAR	&	9182.01259	&	2.43	&	2.01	&	MXB\_20	&	9525.8648	&	14.24	&	0.97	&	MXR\_21	\\
9141.6397	&	-7.51	&	3	&	CAR	&	9182.01259	&	6.32	&	1.21	&	MXR\_20	&	9529.92016	&	-2.64	&	1.38	&	MXB\_21	\\
9141.7027	&	-8.45	&	2.59	&	CAR	&	9182.81351	&	6.12	&	2.15	&	MXB\_20	&	9529.92016	&	-1.32	&	0.84	&	MXR\_21	\\
9142.5187	&	7.29	&	2.74	&	CAR	&	9182.81351	&	1.24	&	1.26	&	MXR\_20	&	9537.98932	&	-2.3	&	1.03	&	MXB\_21	\\
9146.5184	&	-0.12	&	2.91	&	CAR	&	9182.89216	&	0.21	&	1.52	&	MXB\_20	&	9537.98932	&	0.35	&	0.65	&	MXR\_21	\\
9146.6025	&	7.18	&	1.98	&	CAR	&	9182.89216	&	1.05	&	0.94	&	MXR\_20	&	9538.85194	&	0.53	&	1.28	&	MXB\_21	\\
9147.408	&	-3.36	&	2.12	&	CAR	&	9182.97943	&	-1.9	&	1.43	&	MXB\_20	&	9538.85194	&	-1.1	&	0.73	&	MXR\_21	\\
9147.5126	&	-5.02	&	2.86	&	CAR	&	9182.97943	&	-2.5	&	0.88	&	MXR\_20	&	9539.81052	&	-0.02	&	1.29	&	MXB\_21	\\
9149.4108	&	-3.6	&	2.95	&	CAR	&	9183.07999	&	-2.04	&	2.03	&	MXB\_20	&	9539.81052	&	-4.02	&	0.64	&	MXR\_21	\\
9149.5024	&	-5.31	&	2.48	&	CAR	&	9183.07999	&	-3.15	&	1.09	&	MXR\_20	&	9539.85581	&	0.14	&	1.08	&	MXB\_21	\\
9149.5915	&	-6.27	&	2.76	&	CAR	&	9183.76783	&	0.54	&	2.26	&	MXB\_20	&	9539.85581	&	-0.46	&	0.62	&	MXR\_21	\\
9149.6962	&	-3.53	&	3.95	&	CAR	&	9183.76783	&	0.84	&	1.1	&	MXR\_20	&	9539.99341	&	6.26	&	1.54	&	MXB\_21	\\
9150.3895	&	-5.06	&	2.3	&	CAR	&	9183.83016	&	-3.87	&	1.63	&	MXR\_20	&	9539.99341	&	4.67	&	0.91	&	MXR\_21	\\
9151.6239	&	0.51	&	1.78	&	CAR	&	9183.99424	&	2.76	&	1.52	&	MXB\_20	&	9540.05145	&	4.77	&	1.41	&	MXB\_21	\\
9151.7309	&	10.76	&	1.97	&	CAR	&	9183.99424	&	2.5	&	0.9	&	MXR\_20	&	9540.05145	&	5.44	&	0.79	&	MXR\_21	\\
9152.4645	&	7.73	&	1.74	&	CAR	&	9184.07505	&	6.65	&	1.95	&	MXB\_20	&		&		&		&		\\
  \hline
  \hline
  \end{tabular}
\end{center}
\end{table*}


\section{Model parameters and priors for pyaneti fit}

\begin{table*}
\begin{center}
  \caption{Model priors and parameters for Pyaneti fit to the combined RV data set described in \S\ref{sec:rv_analysis}.}  
  \label{tab:pyaneti}
  \begin{tabular}{lcc}
  \hline
  \hline
  \noalign{\smallskip}
  Parameter & Prior$^{(a)}$ & Final value$^{(b)}$ \\
  \noalign{\smallskip}
  \hline
  \noalign{\smallskip}
  \multicolumn{3}{l}{\emph{\bf TOI-1685 b parameters }} \\
  \noalign{\smallskip}
    Orbital period $P_{\mathrm{orb}}$ (days)  & $\mathcal{N}[0.66913923 , 0.00000040]$ & $ 0.66913924 \pm 0.00000042 $ \\
    Transit epoch $T_0$ (BJD$_\mathrm{TDB}-$2\,450\,000)  & $\mathcal{N}[9910.93830 , [0.00038]$ & $ 9910.93828_{-0.00039}^{+0.00041} $  \\  
    Orbital eccentricity, $e$ & $\mathcal{F}[0]$ & 0 \\
    Doppler semi-amplitude variation $K$ (m s$^{-1}$) & $\mathcal{U}[0,50]$ & $ 3.76_{-0.38}^{+0.39} $ \\
    \hline
    \multicolumn{3}{l}{\emph{\bf GP hyperparameters}} \\
    GP Period $P_{\rm GP}$ (days) &  $\mathcal{U}[17,20]$ & $ 18.15_{-0.36}^{+0.39} $ \\
    $\lambda_{\rm p}$ &  $\mathcal{U}[0.01,2]$ &  $ 0.43_{-0.16}^{+0.24} $ \\
    $\lambda_{\rm e}$ (days) &  $\mathcal{U}[1,150]$ & $ 50.8_{-21.0}^{+41.5} $ \\    
    $A_{0}$ &  $\mathcal{U}[0,1.5]$ & $ 0.0066_{-0.0015}^{+0.0032} $ \\
    $A_{1}$ &  $\mathcal{F}[0]$ & 0 \\
    \hline
    \multicolumn{3}{l}{\emph{\bf Other parameters$^{c}$}} \\
    $\sigma_{\mathrm{RV}}$ CARMENES (\ms) & $\mathcal{J}[1,100]$ & $ 3.01_{-0.89}^{+0.71} $ \\
    $\sigma_{\mathrm{RV}}$ IRD (\ms) & $\mathcal{J}[1,100]$ & $ 0.79_{-0.58}^{+1.02} $ \\
    $\sigma_{\mathrm{RV}}$ MAROON-X Blue 2020 (\ms) & $\mathcal{J}[1,100]$ & $ 1.03_{-0.74}^{+0.76} $ \\
    $\sigma_{\mathrm{RV}}$ MAROON-X Red 2020 (\ms) & $\mathcal{J}[1,100]$ & $ 1.37_{-0.43}^{+0.53} $ \\
    $\sigma_{\mathrm{RV}}$ MAROON-X Blue 2021 (\ms) & $\mathcal{J}[1,100]$ & $ 0.86_{-0.62}^{+0.96} $ \\
    $\sigma_{\mathrm{RV}}$ MAROON-X Red 2020 (\ms) & $\mathcal{J}[1,100]$ &  $ 1.09_{-0.59}^{+0.79} $ \\
    $\gamma_{\mathrm{RV}}$ CARMENES (\ms) & $\mathcal{U}[ -5, 5]$ & $ -1.4_{-3.8}^{+3.0} $ \\
    $\gamma_{\mathrm{RV}}$ IRD (\ms) & $\mathcal{U}[ -5, 5]$ & $ -2.1_{-3.6}^{+3.4} $ \\
    $\gamma_{\mathrm{RV}}$ MAROON-X Blue 2020 (\ms) & $\mathcal{U}[ -5, 5]$ & $ -2.1_{4.2}^{+3.7} $ \\
    $\gamma_{\mathrm{RV}}$ MAROON-X Red 2020 (\ms) & $\mathcal{U}[ -5, 5]$ & $ -2.1_{-4.2}^{+3.6} $ \\
    $\gamma_{\mathrm{RV}}$ MAROON-X Blue 2021 (\ms) & $\mathcal{U}[ -5, 5]$ & $ 1.1_{-4.1}^{+4.3} $ \\
    $\gamma_{\mathrm{RV}}$ MAROON-X Red 2020 (\ms) & $\mathcal{U}[ -5, 5]$ & $ 1.1_{-4.1}^{+4.3} $ \\
    \noalign{\smallskip}
    \hline
\multicolumn{3}{l}{\footnotesize $^a$ $\mathcal{F}[a]$ refers to a fixed value $a$, $\mathcal{U}[a,b]$ to an uniform prior between $a$ and $b$, $\mathcal{N}[a,b]$ to a Gaussian prior with mean $a$ and standard deviation $b$,} \\
\multicolumn{3}{l}{\footnotesize and $\mathcal{J}[a,b]$ to the modified Jeffrey's prior as defined by \citet[eq.~16]{Gregory2005}.}\\
\multicolumn{3}{l}{\footnotesize $^b$ Inferred parameters and errors are defined as the median and 68.3\% credible interval of the posterior distribution.}\\
\multicolumn{3}{l}{\footnotesize $^c$ $\sigma_{RV}$ is the instrument- and semester-specific RV jitter value, while $\gamma_{RV}$ is the RV offset applied between the individual data sets.}\\
  \hline
  \hline
  \end{tabular}
\end{center}
\end{table*}

\end{appendix}

\end{document}